\documentclass[%
reprint,
amsmath,amssymb, aps, pra
]{revtex4-2}

\usepackage{dcolumn}
\usepackage{amsmath,amssymb,amsthm,bm,bbm,braket,graphicx,mathtools,mathrsfs,upref,xurl,enumitem,physics,dsfont,multirow,lmodern,rotating,ragged2e,adjustbox,xcolor,comment,hyperref,appendix,scalerel,fix-cm,colortbl,booktabs}
\makeatletter
\newcommand{\shorteq}{%
  \settowidth{\@tempdima}{-}
  \resizebox{\@tempdima}{\height}{=}%
}
\definecolor{grey}{gray}{0.88}
\newcolumntype{a}{>{\columncolor{grey}}c}
\renewcommand{\ge}{\geqslant}
\renewcommand{\le}{\leqslant}

\newtheorem{prop}{Proposition}[section]

\usepackage{soul}
\newcommand{\PB}[1]{\textcolor{blue}{#1}}

\begin{document}


\title{Strength of statistical evidence for genuine tripartite nonlocality}   

\author{Soumyadip Patra}
\author{Peter Bierhorst}%
\affiliation{%
 Department of Mathematics, University of New Orleans, New Orleans, Louisiana 70148, USA
}%




\date{\today}

\begin{abstract}
Recent advancements in network nonlocality have led to the concept of local operations and shared randomness-based genuine multipartite nonlocality (LOSR-GMNL). In this paper, we consider two recent experimental demonstrations of LOSR-GMNL, focusing on a tripartite scenario where the goal is to exhibit correlations impossible in a network where each two-party subset shares bipartite resources and every party has access to unlimited shared randomness. Traditional statistical analyses measuring violations of witnessing inequalities by the number of experimental standard deviations do not account for subtleties such as memory effects. We demonstrate a more sound method based on the prediction-based ratio (PBR) protocol to analyse finite experimental data and quantify the strength of evidence in favour of genuine tripartite nonlocality in terms of a valid $p$-value. In our work, we propose an efficient modification of the test factor optimisation using an approximating polytope approach. By justifying a further restriction to a smaller polytope we enhance practical feasibility while maintaining statistical rigour.  
\end{abstract}

\maketitle

\section{Introduction}\label{s:Intro}

The standard bipartite Bell scenario, where two spatially separated parties perform local measurements on an entangled system, has been pivotal in demonstrating quantum correlations that defy any local hidden variable explanation~\cite{Bell_1964,Bell_1966,CHSH_1969,BrunnerCavalcanti_2014}. Over the past decade, loophole-free experimental demonstrations of such correlations have not only ruled out classical descriptions of Nature~\cite{Hensen2015,Giustina2015,Shalm2015,Rosenfeld2017,Li2018} but have also enabled tasks such as device-independent quantum key distribution~\cite{Ekert91,Acin2007,Pironio2009} and device-independent quantum random number generation~\cite{Pironio2010,Bierhorst2018,Liu2018,Shalm2021}.

The standard notion of Bell nonlocality is insufficient for addressing genuine nonlocality in scenarios involving more than two parties. In a tripartite scenario, according to previously held notions of genuine multipartite nonlocality (GMNL)~\cite{Svetlichny87,Bancal2013}, correlations are considered genuinely multipartite nonlocal if their nonlocality cannot be reduced to bipartite nonlocality; that is, if they cannot be decomposed into convex combinations of exclusively bipartite-nonlocal correlations.

Over the past decade, nonlocality in network scenarios has been extensively researched~\cite{Fritz_2012,Renou_2019,Wolfe_2021,Tavakoli_2022} leading to a new definition, referred to as LOSR-GMNL, based on the framework of local operations and shared randomness (LOSR)~\cite{XCR2021,Coiteux-Roy2021}, which substituted the previously held notion~\cite{Schmid2023} based on local operations and classical communication (LOCC)~\cite{Chitambar2014}. In a tripartite scenario, the new definition can be understood as follows: Consider three spatially separated parties where every two-party subset shares bipartite resources, which may include classical randomness, copies of maximally entangled quantum states, or super-quantum resources like the paradigmatic Popescu-Rohrlich (PR) Box~\cite{PR94}. The parties perform local operations on their portions of the resources they share and have access to unbounded shared local randomness. Correlations inexplicable in such a strategy exhibit LOSR-GMNL.

Recently, genuinely tripartite nonlocal correlations have been experimentally claimed in~\cite{Mao2022,Cao2022,JianWeiPan2022}. In~\cite{Mao2022,Cao2022}, employing the techniques developed in~\cite{WolfeSpekkensFritz_2019,Wolfe_2021}, the authors obtain novel Bell-type inequalities satisfied by tripartite correlations permissible in a strategy based on local operations on bipartite resources supplemented with globally shared local randomness, but violated by appropriate measurements on three-way entangled Greenberger-Horne-Zeilinger (GHZ) states. Then these linear inequalities' violations serve as device-independent (DI) witnesses for genuine tripartite nonlocality (GTNL). Using estimated measurement-settings-conditional outcome probabilities from a large number of experimental trials implementing a photonic tripartite GHZ protocol, both~\cite{Mao2022} and~\cite{Cao2022} claim GTNL by reporting the degree of violation of the DI witnesses in terms of the number of standard deviations (SDs) beyond their maximal value. Both works report a large number of experimental SDs of violation of their respective DI witness.

Although the high degree of violation in terms of the number of SDs reflects the precision with which the DI witnesses are violated, it is not a valid quantification of the strength of statistical evidence for GTNL. This approach fails to properly quantify the probability that, due to statistical fluctuations in a finite number of trials, the DI witnesses could be violated by distributions that still admit an explanation in terms of bipartite resources and shared randomness. Moreover, the method does not account for the potential for a memory attack to have influenced the experimental results~\cite{Barrett2002}. In experiments with a fixed number of trials, the results of a given trial could depend on those from previous trials, especially if the prepared tripartite state and/or measurement settings vary arbitrarily over time.

In this paper, we present a robust method for statistical analysis of experimental data, and quantify the strength of evidence for GTNL. Our method is based on the prediction-based ratio (PBR) protocol, proposed in~\cite{ZhangGlancyKnill2011,ZhangGlancyKnill2013}, where the authors analyse data from Bell-test experiments demonstrating a departure from local realism, and quantify the evidence of departure in terms of a valid $p$-value. Other than the test of local realism and demonstration of GTNL, the PBR protocol is also applicable in statistical analysis of finite data demonstrating other properties such as entanglement detection~\cite{Horodecki96,Terhal2000}, quantification of entanglement~\cite{VidalWerner2002,Toth2015,Arnon-Friedman_2019}, Hilbert space dimension~\cite{BrunnerPironioAcinGisin08}, and fidelity to some target state~\cite{BancalNavascues2015,YangVertesi2014} that involve violating DI linear witnesses such as Bell-type inequalities~\cite{Chang_2024}; furthermore, the PBR protocol is also closely related to probability estimation factors (PEFs) involved in the framework of probability estimation---used in certifying device independent quantum randomness upon finite data obtained from loophole-free Bell experiments~\cite{ZhangKnillBierhorst2018,KnillZhangBierhorst2020,PatraBierhorst2023}. The PBR method allows the possibility that the outcomes in a given trial are dependent on the outcomes and measurement settings of the previous trials, hence the analysis is not subject to the memory loophole. It remains valid in presence of state fluctuations over time, variations in measurement settings and other experimental parameters such as detector efficiency, and is asymptotically optimal~\cite{ZhangGlancyKnill2011}.

In our work, we consider count data from the experimental demonstrations in~\cite{Mao2022,Cao2022}. After implementing a maximum-likelihood-based optimisation procedure to mitigate the presence of weak signalling effects usually present in finite experimental data, we use the derived empirical trial distribution to obtain valid $p$-values quantifying the evidence in favour of GTNL, where the $p$-value is calculated based on a \emph{test statistic} which is a function of the data actually observed in the experiment. The test statistic employed in our work is a trialwise multiplicative accumulation of non-negative real-valued functions of the trial results, which we refer to as \emph{test factors}. The PBR method uses a heuristic-based optimisation routine to find useful test factors by maximising the expected base-2 logarithm of the test factor with respect to the maximum-likelihood no-signalling estimate of the empirical trial distribution, subject to the constraint that the expected value of the test factor is at most one for distributions achievable with bipartite resources and shared randomness. 

Since characterising the set of such distributions remains a challenging problem and the test factor optimisation requires a finite number of constraints, we employ a modification by considering an approximating polytope containing such distributions, obtained from a vertex enumeration routine for the intersection of the tripartite no-signalling polytope (for binary settings and outcomes) and the DI witness. The method of using an approximating polytope for an otherwise hard-to-characterise set has previously been used in the related context of probability estimation\cite{ZhangKnillBierhorst2018, KnillZhangBierhorst2020,BierhorstZhang2020}. A key new aspect of our work is that we are then able to reduce the computation overload by considering instead the polytope comprising no-signalling distributions that saturate the DI witness instead of obeying it, replacing an inequality constraint with an equality constraint, which we justify rigorously. Consequently, we achieved a reduction in the number of optimisation constraints by an order of magnitude. Our reduction technique can be applied to PBR-based statistical analysis for other properties detected by DI linear witnesses, including entanglement detection, Hilbert space dimension, fidelity to a pure state, and probability estimation. This promises to be an important reduction technique as network experiments of increasing numbers of parties, measurement settings, and outcomes require consideration of distribution polytopes of much increased dimension and complexity.

In the next sections, we discuss the specific DI linear witnesses used in~\cite{Mao2022,Cao2022} as signatures for GTNL, the test factor method to quantify the evidence for GTNL, and the analysis of count data from the two experimental demonstrations.

\section{Bell-type test for GTNL}\label{s:Bell_GTNL_witness}

We study a scenario involving three space-like separated parties $\mathsf{A},\mathsf{B},\mathsf{C}$ with respective measurement settings labelled in binary as $x,y,z\in\{0,1\}$, where for each measurement, they observe one out of two possible outcomes also labelled in binary as $a,b,c\in\{0,1\}$. We refer to the settings-conditional outcome probabilities $P(abc|xyz)$ as \textit{behaviours} and, to have a geometric formulation, denote them as vectors $\mathbf{P}\in\mathbb{R}^{64}$, whose components are the $2^6=64$ probabilities $P(abc\lvert xyz)$ of the settings-outcome combinations. The parties are constrained by the relativistic principle of no instantaneous-signalling which means that the behaviours satisfy the no-signalling conditions (see Appendix~\ref{a:NSreview} for a short review of the no-signalling set).

\begin{figure}[!htb]
    \centering
    \includegraphics[width=0.8\linewidth, keepaspectratio]{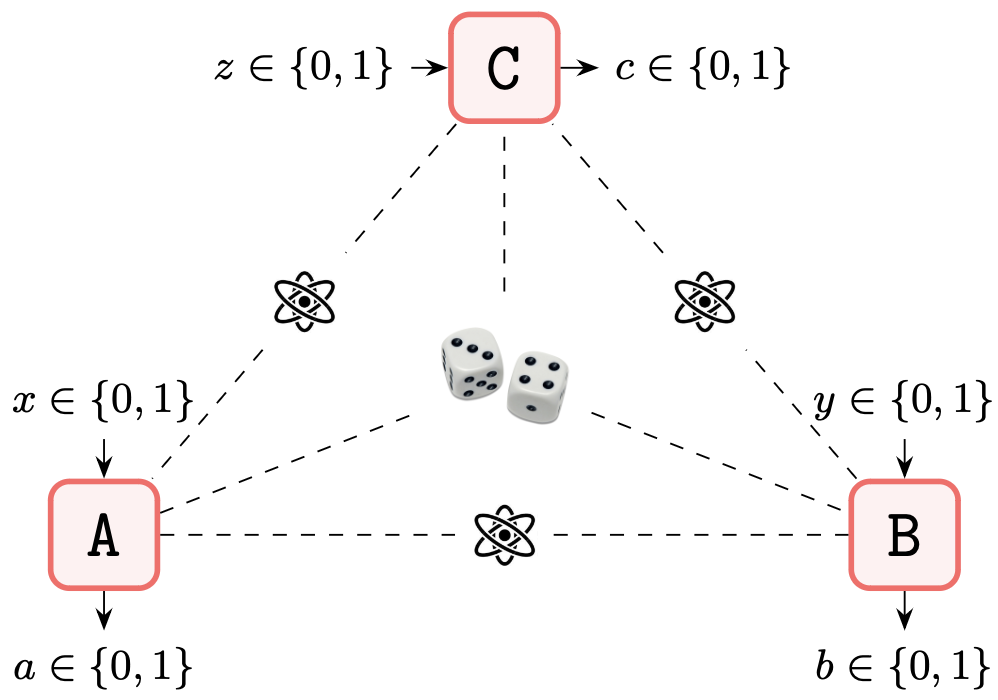}
    \caption{Only behaviours impermissible under a strategy depicted above, involving local operations on (possibly generalised) bipartite nonlocal resources supplemented with unlimited shared randomness, are considered genuinely tripartite nonlocal. The photon symbols shared by every subset of two parties denote the shared bipartite resource. The three-way shared dice represent globally shared local randomness.}
    \label{fig:3partytriangle}
\end{figure}

Figure~\ref{fig:3partytriangle} shows a network of three parties where each subset of two parties shares a (possibly super-quantum) bipartite-nonlocal resource and each party performs local operations on its portion of the bipartite resource shared with the other parties. Additionally, the parties have access to shared local randomness. Bell-type inequalities proposed in~\cite{Mao2022} and~\cite{Cao2022} are obeyed by all behaviours induced by a Figure~\ref{fig:3partytriangle}-style network; behaviours violating these inequalities require three-way nonlocal resources and thus are considered genuinely tripartite nonlocal. In this sense, the inequalities are DI witnesses for GTNL. Both witnesses are obtained theory-agnostically using the inflation techniques developed in~\cite{WolfeSpekkensFritz_2019,Wolfe_2021} under the assumption of compatibility with device replication and a causal structure that includes classical, quantum and generalised probabilistic theories. The witnesses are as shown below:
\begin{eqnarray} \mathcal{M}_{\mathsf{AB}}^{00}+\mathcal{M}_{\mathsf{AB}}^{01} + \mathcal{C}_{\mathsf{ABC}}^{101}-\mathcal{C}_{\mathsf{ABC}}^{111}
    + 2\mathcal{M}_{\mathsf{AC}}^{00}&& \le 4,\label{eq:Mao}\\
\mathcal{M}_{\mathsf{AB}}^{00}-\mathcal{M}_{\mathsf{AB}}^{01} + \mathcal{M}_{\mathsf{BC}}^{00}-\mathcal{M}_{\mathsf{BC}}^{10}+ 2\mathcal{C}_{\mathsf{ABC}}^{101} && \nonumber \\
    + 2\mathcal{C}_{\mathsf{ABC}}^{111} + 4\mathcal{M}_{\mathsf{AC}}^{00}&& \le 8\PB{,}\label{eq:Cao}
\end{eqnarray}
where the marginal correlators ($\mathcal{M}_{\mathsf{A}}^{x}$, $\mathcal{M}_{\mathsf{AB}}^{xy}$, and likewise) and full correlators ($\mathcal{C}_{\mathsf{ABC}}^{xyz}$) are as defined as:
\begin{eqnarray}
    \mathcal{M}_{\mathsf{A}}^{x} &\coloneqq& \sum_{a}(-1)^{a}P(a\lvert x)\label{eq:MarginCorrA} \\
    \mathcal{M}_{\mathsf{AB}}^{xy} &\coloneqq& \sum_{a,b}(-1)^{a\oplus b}P(ab\lvert xy)\label{eq:MarginCorrAB}\\
    \mathcal{C}_{\mathsf{ABC}}^{xyz} &\coloneqq& \sum_{a,b,c}(-1)^{a\oplus b\oplus c}P(abc\lvert xyz).\label{eq:FullCorr}
\end{eqnarray}
Interestingly, the witness in~\eqref{eq:Cao} can be derived from that in~\eqref{eq:Mao}, as demonstrated in~\cite{Bierhorst2024}, by adding together two relabelled versions of~\eqref{eq:Mao}. However, as pointed out in~\cite{Bierhorst2024} that since the two relabelled versions of~\eqref{eq:Mao} are each linear witnesses themselves, i.e., their violation requires genuinely tripartite nonlocal behaviours, their sum which is~\eqref{eq:Cao} is not necessarily a weaker witness of GTNL than~\eqref{eq:Mao}. In fact, in section~\ref{s:Data_Analysis} we infer that it is the contrary for a specific case involving the depolarised GHZ behaviour. The fact that the two witnesses are inequivalent, even though one is derivable from the other, will be more evident later in section~\ref{s:Data_Analysis} where we compare the two polytopes obtained by the intersection of the no-signalling set with the corresponding linear witnesses for our data analyses.

The maximum quantum violations, $2+2\sqrt{2}$ and $4+4\sqrt{2}$ for witnesses~\eqref{eq:Mao} and~\eqref{eq:Cao}, respectively, are achieved by behaviours realisable in a protocol involving the pure state $\ket{\mathsf{GHZ}}=\frac{1}{\sqrt{2}}(\ket{000}+\ket{111})$ shared by the three parties where each party performs localised projective measurements $\mathbf{r}\cdot\boldsymbol{\sigma}$ in suitable directions $\mathbf{r}\in\mathbb{S}^{2}$ (see the ideal protocol described in~\cite{Mao2022} and~\cite{Cao2022} for the directions). The vector $\boldsymbol{\sigma}\equiv(\mathsf {X,Y,Z})$ represents the vector of $2\times 2$ Pauli matrices. Using the linear relations in~\eqref{eq:CorrVector} we can express the inequalities in~\eqref{eq:Mao} and~\eqref{eq:Cao} as 
\begin{equation*}
 \sum_{a,b,c,x,y,z}B(abcxyz)P(abc\lvert xyz)\le\beta,   
\end{equation*}
which we abbreviate using a geometric formulation as $\mathbf{B}\cdot\mathbf{P}\le\beta$, where $\mathbf{B}\in\mathbb{R}^{64}$ is a Bell vector of the $B(abcxyz)$ coefficients. Note the marginal correlators appearing in the DI witnesses in ~\eqref{eq:Mao} and~\eqref{eq:Cao} are well-defined due to the no-signalling conditions, depending only on the measurement settings of the parties involved. For instance, $\mathcal{M}_{\mathsf A}^{x}=\mathcal{M}_{\mathsf A}^{xyz}$ and 
$\mathcal{M}_{\mathsf{AB}}^{xy}=\mathcal{M}_{\mathsf{AB}}^{xyz}$. Consequently, the Bell vectors corresponding to the witnesses are not unique. Tables~\ref{tab:B1} and~\ref{tab:B2} in Appendix~\ref{a:VertEnum} show two possible Bell vectors for the witnesses, $\mathbf{B}_{(1)}$ for~\eqref{eq:Mao} and $\mathbf{B}_{(2)}$ for~\eqref{eq:Cao}.

\section{Strength of statistical evidence using test factors}\label{s:Strength_Of_Evidence}

An experiment demonstrating GTNL consists of a number of trials. Each trial is associated with outcomes and measurement settings collectively referred to as the trial results. The trial results are modelled as a random tuple $(A_{k},B_{k},C_{k},X_{k},Y_{k},Z_{k})$ with $k\in[n]$. The \textit{experiment results} are then denoted by the sequence of tuples $\{(A_{k},B_{k},C_{k},X_{k},Y_{k},Z_{k})\}_{k=1}^{n}$, where $n$ is the number of trials. We are interested in quantifying the strength of statistical evidence, by means of a $p$-value, against the hypothesis $\mathcal{H}_{\mathsf{lo2sr}}$ described as follows: \textit{The trial results follow a distribution permissible within a strategy encompassing local operations on (possibly generalised) bipartite-nonlocal resources and unlimited shared local randomness}. Experiment results inconsistent with a behaviour satisfying \eqref{eq:Mao} and \eqref{eq:Cao} provide evidence against this hypothesis.

A $p$-value can be defined in association with a function $T$ of the experiment results, referred to as a test statistic, as
\begin{equation}\label{eq:p_value}
    p = \sup_{\mathsf{lo2sr}}\mathsf{Prob}[T(\mathbf{U}_{\mathsf{lo2sr}})\ge T(\mathbf{u})].
\end{equation}
In~\eqref{eq:p_value}, $\mathbf{U}_{\mathsf{lo2sr}}$ is a sequence $U_{1},U_{2},\ldots,U_{n}$, where $U_{k}$ is the random tuple $(A_{k},B_{k},C_{k},X_{k},Y_{k},Z_{k})$ following a distribution consistent with $\mathcal{H}_{\mathsf{lo2sr}}$; the supremum is taken over all such distributions. The test statistic is composed of a product of non-negative functions of trial results $F(U_{k})$ which we refer to as test factors and whose expected value is at most one for $U_{k}$ distributed according to $\mathcal{H}_{\mathsf{lo2sr}}$, so that test factors exceeding one on average suggest inconsistency with $\mathcal{H}_{\mathsf{lo2sr}}$. The test factor $F$ can in general be updated from trial to trial or from a batch of trials to another batch  during run time~\cite{ZhangGlancyKnill2011}, but for this work we consider a fixed function for all trials.

The test statistic is then defined as a trial-wise product of the test factors, i.e., $T(u_{1},\ldots,u_{n})\coloneqq\prod_{k=1}^{n}F(u_{k})$. The following formula for $p_{(\mathsf{PBR})}$ is a valid upper bound for the exact $p$-value (defined as the maximal probability that the test statistic is at least as extreme as the value actually observed from the data under the assumption that the null hypothesis is true):
\begin{equation}\label{eq:p_val}
p_{(\mathsf{PBR})} = \min\left\{\Big(\prod_{k=1}^{n}F(u_{k})\Big)^{-1},1\right\}. \end{equation}
Validity of this p-value bound has been shown in~\cite{ZhangGlancyKnill2011} (refer to the arguments between (16) and (17) in section III~C) where the null hypothesis is Local Realism, and can be adapted straigthforwardly for the null hypothesis considered in our work, i.e., $\mathcal{H}_{\mathsf{lo2sr}}$. Henceforth, we will refer to the upper bound in~\eqref{eq:p_val} to the exact $p$-value as a valid $p$-value.

Our method of obtaining useful test factors is to seek a high expected value of $\frac{1}{n}\log_{2}T(U_{1},\ldots,U_{n})=\frac{1}{n}\sum_{i=1}^{n}\log_{2}F(U_{i})$~\cite{ZhangGlancyKnill2011}. Heuristically, in constructing the test factor $F$ we assume a stable experiment with independent trials, where future results will be identically distributed according to an estimate of the true distribution typically derived from a reserved portion of the data from previous trials. Now, whether or not such assumptions are actually met only affects the quality of the $p$-value, not its validity. For the remainder of this paper a trial distribution $\mathbb{P}\coloneqq\{P(abc\lvert xyz)S(xyz)\}$ is associated with a trial behaviour $\mathbf{P}$ assuming a known and fixed settings distribution $\{S(xyz)\}$; hence the term \textit{distribution} in this context refers to unconditional joint probability distribution of trials which can be obtained from the associated \textit{behaviour} (settings-condition outcome distribution) by multiplying by the settings distribution. We then perform the following optimisation to obtain useful TFs:
\begin{eqnarray}\label{eq:TestFactor_Opt}
    \max_{F}\,\, &&\mathsf{E}_{\mathbb{Q}}[\log_{2}F(ABCXYZ)] \nonumber \\
    \text{subject to}\,\, &&\mathsf{E}_{\mathbb{P}}[F(ABCXYZ)]\le 1,\, \forall \mathbf{P}\in\Delta_{\mathsf{lo2sr}}, \nonumber \\
    &&F(abcxyz)\ge 0, \,\forall a,b,c,x,y,z,
\end{eqnarray}
where $\mathbb{Q}$ is the anticipated trial distribution assuming a stable experiment and $\Delta_{\mathsf{lo2sr}}$ is the set of behaviours consistent with $\mathcal{H}_{\mathsf{lo2sr}}$. The idea behind the objective function is that based on our heuristic, for large values of $n$, the difference between the observed value of $\frac{1}{n}\sum_{k=1}^{n}\log_{2}F(U_{k})$ and the expected value of $\log_{2}F(U)$ will be greater or less than zero with roughly equal probability due to the central limit theorem. And so the test factor achieving the optimum value will tend to yield the largest observed product of test factors. The objective in~\eqref{eq:TestFactor_Opt} always returns a non-negative value as the test factor defined as $F(abcxyz)=1,\,\forall a,b,c,x,y,z$, is valid and its logarithm has an expected value of zero for any distribution. A positive expected $\log_{2}F_{\mathsf{opt}}$ with respect to the anticipated distribution $\mathbb{Q}$, where $F_{\mathsf{opt}}$ is the optimal test factor according to the optimisation, can be expected if $\mathbb{Q}$ is not consistent with $\mathcal{H}_{\mathsf{lo2sr}}$. 

The set $\Delta_{\mathsf{lo2sr}}$ is convex---the allowance of globally shared randomness in the model ensures that convex mixtures of behaviours in $\Delta_{\mathsf{lo2sr}}$ are again in $\Delta_{\mathsf{lo2sr}}$---and hence, linearity of expectation justifies restricting the constraint $\mathsf{E}_{\mathbb{P}}[F]\le 1$ to the set of extreme points of $\Delta_{\mathsf{lo2sr}}$, as this will imply it holds for all behaviours in the convex set. However, a complete characterisation of $\Delta_{\mathsf{lo2sr}}$ in terms of its extreme points remains a challenging open question. Furthermore, it is possible that it has curved boundaries in some regions, which would imply a continuum of extreme points. The test factor method, on the other hand, requires a finite set of optimisation constraints for the problem to be computable, i.e., in~\eqref{eq:TestFactor_Opt} the set of behaviours $\mathbf{P}$ for which $\mathsf{E}_{\mathbb P}[F]\le 1$ holds, must be extreme points of a polytope containing $\Delta_{\mathsf{lo2sr}}$. In the forthcoming data analysis we demonstrate how we address this issue by making some modifications to the feasibility region of~\eqref{eq:TestFactor_Opt}.

\section{Data analysis}\label{s:Data_Analysis}

\subsection{Experimental data from Mao et al~\texorpdfstring{\cite{Mao2022}}{Bb}}

First, we present the analysis of count data for the experimental demonstration in~\cite{Mao2022} obtained from~\cite{PrivateComm}. An empirical trial distribution is derived from the experimental data as $f_{\mathsf{emp}}(abcxyz)=\frac{N(abc\lvert xyz)}{\sum_{abc}N(abc\lvert xyz)}S(xyz)$, where $N(abc\lvert xyz)$ represents the count of outcome combinations $a,b,c$ given the setting combinations $x,y,z$, and $S(xyz)$ is the fixed settings distribution which we take to be uniform, i.e., $S(xyz)=1/8,\,\forall x,y,z$. Due to statistical fluctuations the empirical distribution might not satisfy the no-signalling conditions exactly. We therefore begin by obtaining a maximum-likelihood no-signalling (MLNS) estimate $\hat{\mathbb Q}$ of the empirical trial distribution as described in Appendix~\ref{a:MLNSestimate} and subsection~\ref{a:MLNS_Mao}. Next, using this estimate we perform the following test factor optimisation:
\begin{eqnarray}\label{eq:TFoptMao}
    \max_{F}\,\, &&\mathsf{E}_{\hat{\mathbb{Q}}}[\log_{2}F(ABCXYZ)] \nonumber \\
    \text{subject to}\,\, &&\mathsf{E}_{\mathbb{P}}[F(ABCXYZ)]\le 1,\, \forall \mathbf{P}\in\mathsf{Ext}(\Xi_{(1)}'), \nonumber \\
    &&F(abcxyz)\ge 0, \,\forall a,b,c,x,y,z.
\end{eqnarray}
Notice in~\eqref{eq:TFoptMao} that the condition $\mathsf{E}_{\mathbb P}[F]\le 1$ is with respect to distributions $\mathbb P$ whose corresponding behaviours $\mathbf{P}$ belong to $\mathsf{Ext}(\Xi_{(1)}')$, a set different from $\Delta_{\mathsf{lo2sr}}$ as appears in \eqref{eq:TestFactor_Opt}. As discussed in the paragraph following~\eqref{eq:TestFactor_Opt}, this is because we need to replace $\Delta_{\mathsf{lo2sr}}$ with an approximating polytope to make the optimisation computable. A suitable candidate would be the set $\Xi_{(1)}\coloneqq\{\mathbf{P}\in\Xi_{\mathsf{ns}}\colon\mathbf{B}_{(1)}\cdot\mathbf{P}\le 4\}$, i.e., the set of behaviours belonging to the tripartite no-signalling polytope $\Xi_{\mathsf{ns}}$ (for binary inputs and outcomes) and satisfying the inequality in~\eqref{eq:Mao}. This results in a polytope lying in a $26$-dimensional affine subspace of the ambient space $\mathbb{R}^{64}$. Due to the linearity of expectation and convexity of $\Xi_{(1)}$ we can further restrict the condition $\mathsf{E}_{\mathbb P}[F]\le 1$ to only those distributions $\mathbb P$ whose corresponding behaviours $\mathbf P$ are the extreme points of $\Xi_{(1)}$. We denote the set of extreme points of $\Xi_{(1)}$ by $\mathsf{Ext}(\Xi_{(1)})$. Listing the extreme points, or \textit{vertices}, of a polytope expressed as an intersection of finite number of hyperplanes and closed half-spaces is known as the \textit{Vertex Enumeration} problem (refer to Appendix~\ref{a:VertEnum} and Table~\ref{tab:VertEnum} for a summary of the vertex enumeration routines considered in this work). Now, the cardinality of $\mathsf{Ext}(\Xi_{(1)})$ corresponds to the number of optimisation constraints in~\eqref{eq:TFoptMao}, and $|\mathsf{Ext}(\Xi_{(1)})|=56767$.

We were able to reduce this number of optimisation constraints from $56767$ to $3200$, thereby streamlining the computation due to a reduction of more than an order of magnitude, by replacing the set $\Xi_{(1)}$ with the $25$-dimensional polytope described as $\Xi_{(1)}'\coloneqq \{\mathbf{P}\in\Xi_{\mathsf{ns}}\colon\mathbf{B}_{(1)}\cdot\mathbf{P}=4\}$. This polytope contains behaviours belonging to $\Xi_{\mathsf{ns}}$ that saturate~\eqref{eq:Mao}. The number of extreme points were found to be $|\mathsf{Ext}(\Xi_{(1)}')|=3200$. We justify this reduction by showing that if the optimisation over $\Xi_{(1)}'$ is performed for a $\hat{\mathbb Q}$ corresponding to a behaviour $\mathbf Q$ for which $\mathbf B_{(1)} \cdot \mathbf Q >4$ (and we would only be interested in running the optimisation in such a case), then the resulting optimising test factor $F_{\mathsf{opt}}$ will not just satisfy $\mathsf{E}_{\mathbb{P}_{\mathsf{sat}}}[F_{\mathsf{opt}}]\le 1$ for $\mathbb{P}_{\mathsf{sat}}$ whose corresponding (no-signalling) behaviour $\mathbf{P}_{\mathsf{sat}}$ saturates the witness in~\eqref{eq:Mao}, but will also satisfy $\mathsf{E}_{\mathbb{P}_{\mathsf{less}}}[F_{\mathsf{opt}}]\le 1$ for all $\mathbb{P}_{\mathsf{less}}$ whose corresponding (no-signalling) behaviour $\mathbf{P}_{\mathsf{less}}$ strictly satisfies the same witness. This method of reducing the number of constraints in the test factor optimisation will be applicable more generally in scenarios where statistical evidence is sought against a theory (modelled as a null hypothesis, for instance, $\mathcal{H}_{\mathsf{lo2sr}}$ in this work) by means of a test employing linear witnesses and the prediction-based ratio (PBR) method. 

\begin{figure}[!htb]
    \centering
    \includegraphics[width=0.8\linewidth,keepaspectratio]{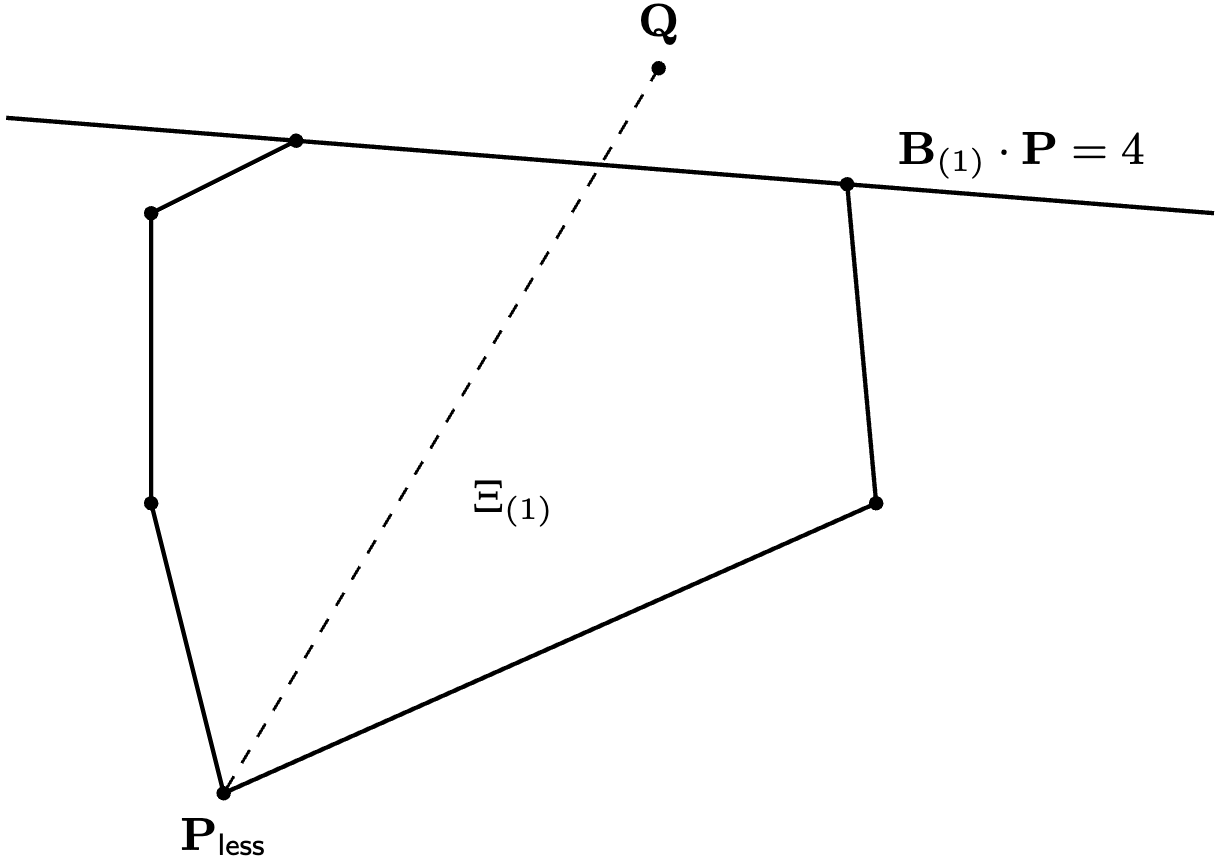}
    \caption{A schematic diagram aiding in our proof for $\mathsf{E}_{\mathbb{P}_{\mathsf{sat}}}[F]\le 1\Rightarrow \mathsf{E}_{\mathbb{P}_{\mathsf{less}}}[F]\le 1$. The irregular polygon represents the $26$-dimensional polytope $\Xi_{(1)}$ resulting from the intersection $\Xi_{\mathsf{ns}}$ and the halfspace $\mathbf{B}_{(1)}\cdot\mathbf{P}\le4$. The points $\mathbf{Q}$ and $\mathbf{P}_{\mathsf{less}}$, respectively, violate and strictly satisfy the inequality $\mathbf{B}_{(1)}\cdot\mathbf{P}\le 4$. The portion of the hyperplane touching $\Xi_{(1)}$ represent a $25$-dimensional polytope. The point $\mathbf{P}_{\mathsf{sat}}$ described in the main text is located at the intersection of the hyperplane and the dashed line.}
    \label{fig:Intersect_NS_Bell}
\end{figure}

We now prove the implication as follows. Consider the schematic diagram shown in Figure~\ref{fig:Intersect_NS_Bell}. The irregular polygon represents the $26$-dimensional polytope $\Xi_{(1)}$ and the line touching it is the ($25$-dimensional) Bell hyperplane $\mathbf{B}_{(1)}\cdot\mathbf{P}=4$. No-signalling behaviours lying in the intersection of the hyperplane and $\Xi_{(1)}$ are the ones that saturate the inequality in~\eqref{eq:Mao}. Next, consider a point $\mathbf{P}_{\mathsf{less}}$ strictly satisfying $\mathbf{B}_{(1)}\cdot\mathbf{P}\le 4$, hence lying below the hyperplane, and the point $\mathbf{}\mathbf{Q}$ (corresponding to the distribution $\hat{\mathbb Q}$ in the objective function of \eqref{eq:TFoptMao}) strictly violating the inequality, hence lying above the hyperplane. The points are labelled in the diagram. The dashed line joining the two points represents all points expressible as the convex combination $\lambda\mathbf{Q}+(1-\lambda)\mathbf{P}_{\mathsf{less}}$ for $\lambda\in[0,1]$. Because the points $\mathbf{Q}$ and $\mathbf{P}_{\mathsf{less}}$ satisfy, respectively, $\mathbf{B}_{(1)}\cdot\mathbf{Q}>4$ and $\mathbf{B}_{(1)}\cdot\mathbf{P}_{\mathsf{less}}<4$, there is some $\lambda\in(0,1)$ for which $\mathbf{B}_{(1)}\cdot(\lambda\mathbf{Q}+(1-\lambda)\mathbf{P}_{\mathsf{less}})=4$, i.e., for that specific value of $\lambda$ the point $\lambda\mathbf{Q}+(1-\lambda)\mathbf{P}_{\mathsf{less}}$ lies in the intersection of the Bell hyperplane and the polytope; let us denote this point as $\mathbf{P}_{\mathsf{sat}}=\lambda\mathbf{Q}+(1-\lambda)\mathbf{P}_{\mathsf{less}}$. We have $\mathsf{E}_{\mathbb{P}_{\mathsf{sat}}}[F_{\mathsf{opt}}]\le 1$, as the optimisation was performed to enforce this constraint for such distributions. Then expressing $\mathbf{P}_{\mathsf{less}}$ in terms of $\mathbf{P}_{\mathsf{sat}}$ and $\mathbf{Q}$ as $\mathbf{P}_{\mathsf{less}}=\frac{1}{1-\lambda}(\mathbf{P}_{\mathsf{sat}}-\lambda\mathbf{Q})$ we can show that the condition $\mathsf{E}_{\mathbb{P}_{\mathsf{sat}}}[F_{\mathsf{opt}}]\le 1$ implies $\mathsf{E}_{\mathbb{P}_{\mathsf{less}}}[F_{\mathsf{opt}}]\le 1$ ( $\mathbb{P}_{\mathsf{less}}$ and $\mathbb{P}_{\mathsf{sat}}$ are the distributions corresponding to the behaviours $\mathbf{P}_{\mathsf{less}}$ and $\mathbf{P}_{\mathsf{sat}}$): By the linearity of expectation we have $\mathsf{E}_{\mathbb{P}_{\mathsf{less}}}[F_{\mathsf{opt}}]=\frac{1}{1-\lambda}\left(\mathsf{E}_{\mathbb{P}_{\mathsf{sat}}}[F_{\mathsf{opt}}]-\lambda\mathsf{E}_{\hat{\mathbb Q}}[F_{\mathsf{opt}}]\right)$. As noted in the discussion following \eqref{eq:TestFactor_Opt}, $\mathsf{E}_{\hat{\mathbb Q}}[\log_{2}F_{\mathsf{opt}}]\ge 0$ will always hold. Then, as $\log_{2}\left(\mathsf{E}_{\hat{\mathbb Q}}[F_{\mathsf{opt}}]\right)\ge\mathsf{E}_{\hat{\mathbb Q}}[\log_{2}F_{\mathsf{opt}}]$ holds (due to Jensen's inequality) we can conclude that $\mathsf{E}_{\hat{\mathbb Q}}[F_{\mathsf{opt}}]\ge 1$ which implies $-\lambda\mathsf{E}_{\hat{\mathbb Q}}[F_{\mathsf{opt}}]\le-\lambda$. Going back to the expression $\mathsf{E}_{\mathbb{P}_{\mathsf{less}}}[F_{\mathsf{opt}}]=\frac{1}{1-\lambda}\left(\mathsf{E}_{\mathbb{P}_{\mathsf{sat}}}[F_{\mathsf{opt}}]-\lambda\mathsf{E}_{\hat{\mathbb{Q}}}[F_{\mathsf{opt}}]\right)$, we then deduce that $\mathsf{E}_{\mathbb{P}_{\mathsf{less}}}[F_{\mathsf{opt}}]\le\frac{1}{1-\lambda}\left(\mathsf{E}_{\mathbb{P}_{\mathsf{sat}}}[F_{\mathsf{opt}}]-\lambda\right)\le 1$. Therefore, it suffices to have the test factor optimisation constraint $\mathsf{E}[F]\le 1$ with respect to only those distributions whose associated behaviours belong to $\Xi_{(1)}'$. 

Subsequently, performing the optimisation in~\eqref{eq:TFoptMao} with respect to the MLNS estimate $\hat{\mathbb Q}$ shown in Table~\ref{tab:MLNSest_Mao_1} in Appendix~\ref{a:MLNSestimate}, we obtain the optimal value $\mathsf{E}_{\hat{\mathbb Q}}[\log_{2}F_{\mathsf{opt}}]\approx 0.017984$. The optimal test factor $F_{\mathsf{opt}}$ is presented in Table~\ref{tab:OptimalTF_Mao}. The evidence in favour of GTNL is quantified by the $p$-value which can be computed using $F_{\mathsf{opt}}$ and the count data from the experiment, shown in~\ref{tab:EmpiricalDist_Mao_1}. From the formula given in~\eqref{eq:p_val} a valid $p$-value can be obtained by inverting the product $\prod_{k=1}^{n}F_{\mathsf{opt}}(u_{k})$, where $u_{k}$ denotes the trial results $(a_k,b_k,c_k,x_k,y_k,z_k)$ for the $k$'th trial. That is, we find the product (over all setting-outcome combinations) of $F_{\mathsf{opt}}(u)$ raised to the power of the corresponding count $\mathsf{ct}(u)$ obtained from the count data in Table~\ref{tab:EmpiricalDist_Mao_1} and invert it, as shown below:
\begin{equation*}
    p = \prod_{u}\Big(F_{\mathsf{opt}}(u)^{\mathsf{ct}(u)}\Big)^{-1} \approx 3.691\times 10^{-71}. 
\end{equation*}
The extremely small $p$-value for this count data, which we will also notice for the other data, is not surprising given that the experimental violation of $4.6674\pm 0.0323$ of the witness in~\eqref{eq:Mao} reported in~\cite{Mao2022} is more than $20$ standard deviations beyond the mean value of $4$. Notice that the $p$-values obtained in practice will be worse than the ones we obtained in our work. This is because in practice a portion of experimental data is set aside for testing, and the optimal test factor $F_{\mathsf{opt}}$ obtained from the training data is used on the test data to calculate a $p$-value. The only data we had at our disposal was the count data for both the experimental demonstrations in~\cite{Mao2022} and~\cite{Cao2022}, and we use $F_{\mathsf{opt}}$ on the same data to calculate a $p$-value from which we obtained it. A test statistic optimised for a data set and then retroactively applied to the same data will in general suggest overly optimistic statistical evidence. However, the $p$-value we obtained is a good rough estimate for what one can expect to find in practice.

\begin{table*}[!htb]
\caption{\label{tab:OptimalTF_Mao}Optimal test factor $F_{\mathsf{opt}}$ obtained from the optimisation in~\eqref{eq:TFoptMao} with respect to the maximum-likelihood no-signalling estimate~\eqref{tab:MLNSest_Mao_1} of the empirical trial distribution. We use uniform settings distribution $S(xyz)=1/8,\,\forall x,y,z$.}
\resizebox{0.95\textwidth}{!}{%
\begin{ruledtabular}
\begin{tabular}{rrcccccccc}
\multicolumn{1}{r}{} & \multicolumn{1}{r}{} & \multicolumn{8}{c}{$abc$}\\
\cline{2-10}
\multicolumn{1}{r}{} & \multicolumn{1}{r}{} & \multicolumn{1}{c}{$000$} & \multicolumn{1}{c}{$001$} & \multicolumn{1}{c}{$010$} & \multicolumn{1}{c}{$011$} & \multicolumn{1}{c}{$100$} & \multicolumn{1}{c}{$101$} & \multicolumn{1}{c}{$110$} & \multicolumn{1}{c}{$111$} \\
\cline{2-10}
\multirow{8}{*}{\rotatebox[origin=c]{90}{$xyz$}} & $000$ & 1.0598122 & 0.6337174 & 0.8904132 & 0.1605416 & 0.1597108 & 0.8806786 & 0.6151026 & 1.0322941\\
 & $001$ & 1.0603007 & 1.0545397 & 0.8396764 & 0.8350988 & 0.7859199 & 0.7957258 & 1.0224645 & 1.0334538\\
 & $010$ & 1.0451923 & 0.6765703 & 0.8714717 & 0.211803 & 0.2503016 & 0.8985366 & 0.6925647 & 1.0497528\\
 & $011$ & 1.0404143 & 1.0480199 & 0.7723418 & 0.7959361 & 0.8458675 & 0.8379063 & 1.0437713 & 1.0517986\\
 & $100$ & 1.0331707 & 0.8675255 & 0.879014 & 1.0171452 & 1.0395271 & 0.8790457 & 0.875241 & 1.018536\\
 & $101$ & 1.17889 & 0.6058302 & 0.64111 & 1.1488267 & 0.5833145 & 1.1724989 & 1.1376239 & 0.6436649\\
 & $110$ & 1.0152896 & 0.8936919 & 0.8646207 & 1.0340698 & 1.0271972 & 0.9004357 & 0.8652325 & 1.0295178\\
 & $111$ & 0.5669301 & 1.1381454 & 1.1593922 & 0.6326589 & 1.1442416 & 0.5532124 & 0.6660395 & 1.1409818\\ 
\end{tabular}
\end{ruledtabular}}
\end{table*}

A potential issue with analysing the raw count data from experiments demonstrating GTNL is the presence of zero frequencies for certain settings-outcomes combinations. When there are zero-valued entries in the empirical trial distribution, the PBR method may assign a zero value to the test factor score for the corresponding settings-outcome combinations. If these combinations occur in a subsequent trial, the product in the expression for $p_{(\mathsf{PBR})}$ in~\eqref{eq:p_val} will be set to zero, which will result in $p_{(\mathsf{PBR})}$ being set to $1$, without any possibility for later adjustment.  Another count data set for the demonstration in~\cite{Mao2022} obtained from~\cite{PrivateComm} that we analyse is as shown in Table~\ref{tab:Rawdata_Mao_2} where the derived three-party data has some instances of zero counts. The analysis for this data is same as before, except that after obtaining the MLNS estimate using the optimisation routine in~\eqref{eq:MLNSestimate} with respect to the empirical trial distribution, we mix the estimate with some noise to address the zero count issue. Using this noise-mixed estimate we then perform the test factor optimisation. We explain these steps in more detail in Appendix~\ref{a:MLNS_Mao}.

\subsection{Experimental data from Cao et al~\texorpdfstring{\cite{Cao2022}}{Cc}}

Next, we analyse count data from the experimental demonstration in~\cite{Cao2022} (as reported in Table I in Appendix A), and reproduced here in Table~\ref{tab:RawData_Cao} in Appendix~\ref{a:MLNS_Cao}. The data in the table pertains to a four-party experiment from which we derive the relevant three-party data by following similar steps as in our previous analysis. During this process, we encountered a missing data issue for two settings combinations, specifically $xyz\equiv000,010$. We addressed this problem by utilising existing data (refer to the row corresponding to the measurement combination $\mathsf{ZZZX}$ in Table~\ref{tab:RawData_Cao}) from the original four-party experiment, along with a no-signalling argument. This allowed us to compute a MLNS estimate with a slight modification in the optimisation procedure; detailed explanations are provided in Appendix~\ref{a:MLNS_Cao}. 
\begin{table*}[!htb]
\caption{\label{tab:OptimalTF_Cao}Optimal test factor $F_{\mathsf{opt}}$ obtained from the optimisation in~\eqref{eq:TFoptCao} with respect to the maximum-likelihood no-signalling estimate~\eqref{tab:MLNSest_Cao} of the empirical trial distribution. We use uniform settings distribution $S(xyz)=1/8,\,\forall x,y,z$.}
\resizebox{0.95\textwidth}{!}{%
\begin{ruledtabular}
\begin{tabular}{rrcccccccc}
\multicolumn{1}{r}{} & \multicolumn{1}{r}{} & \multicolumn{8}{c}{$abc$}\\
\cline{2-10}
\multicolumn{1}{r}{} & \multicolumn{1}{r}{} & \multicolumn{1}{c}{$000$} & \multicolumn{1}{c}{$001$} & \multicolumn{1}{c}{$010$} & \multicolumn{1}{c}{$011$} & \multicolumn{1}{c}{$100$} & \multicolumn{1}{c}{$101$} & \multicolumn{1}{c}{$110$} & \multicolumn{1}{c}{$111$} \\
\cline{2-10}
\multirow{8}{*}{\rotatebox[origin=c]{90}{$xyz$}} & $000$ & 1.0073726 & 0.3733147 & 1.0073726 & 0.3733147 & 0.3679765 & 1.0122465 & 0.3679765 & 1.0122465 \\
 & $001$ & 1.0571756 & 1.0592063 & 0.745292 & 0.772362 & 0.7352839 & 0.7331479 & 1.0625527
 & 1.0854559 \\
 & $010$ & 1.0073726 & 0.3733813 & 1.0073726 & 0.3733813 & 0.3679563 & 1.0122465 & 0.3679563
 & 1.0122465 \\
 & $011$ & 0.7481449 & 0.7434019 & 1.0714002 & 1.0585814 & 1.0705442 & 1.069968 & 0.7546472
 & 0.7459951 \\
 & $100$ & 1.0638622 & 0.7369682 & 0.7533695 & 1.0656278 & 1.0529862 & 0.7196825 & 0.7692114
 & 1.0750601 \\
 & $101$ & 1.1853197 & 0.5666988 & 0.5293914 & 1.1640358 & 0.5684837 & 1.1670996 & 1.1641421
 & 0.4594142 \\
 & $110$ & 0.7469697 & 1.0636082 & 1.0670628 & 0.744549 & 0.7476439 & 1.070692 & 1.0626543
 & 0.7465501 \\
 & $111$ & 1.1837828 & 0.5268113 & 0.5404343 & 1.1698433 & 0.5325159 & 1.2149169 & 1.1725549
 & 0.5847271 \\
\end{tabular}
\end{ruledtabular}}
\end{table*}

The intersection of the polytope $\Xi_{\mathsf{ns}}$ with the witness~\eqref{eq:Cao} (equivalently, $\mathbf{B}_{(2)}\cdot\mathbf{P}\le 8$) results in the $26$-dimensional polytope $\Xi_{(2)}$, where $|\mathsf{Ext}(\Xi_{(2)})|=57283$. Notice the difference in cardinality from our previous polytope ($|\mathsf{Ext}(\Xi_{(1)})|=56767$) provides evidence that \eqref{eq:Mao} cannot be derived from \eqref{eq:Cao} through relabelling, which are symmetries of the no-signalling polytope, or through no-signalling adjustments to the inequalities. The $25$-dimensional polytope $\Xi_{(2)}'$ comprises no-signalling behaviours that saturate~\eqref{eq:Cao} (equivalently, satisfy $\mathbf{B}_{(2)}\cdot\mathbf{P}=8$), and $|\mathsf{Ext}(\Xi_{(2)}')|=3664$. The restriction of the optimisation feasibility region by means of having the condition $\mathsf{E}[F]\le 1$ with respect to only those distributions whose corresponding behaviours belong to the set $\mathsf{Ext}(\Xi_{(2)}')$ is based on the same reasoning as in our previous data analysis. 

After obtaining a MLNS estimate, we proceed to the test factor optimisation as shown below.
\begin{eqnarray}\label{eq:TFoptCao}
    \max_{F}\,\, &&\mathsf{E}_{\hat{\mathbb{Q}}}[\log_{2}F(ABCXYZ)] \nonumber \\
    \text{subject to}\,\, &&\mathsf{E}_{\mathbb{P}}[F(ABCXYZ)]\le 1,\, \forall \mathbf{P}\in\mathsf{Ext}(\Xi_{(2)}'), \nonumber \\
    &&F(a0c000)=F(a1c000),\,\forall a,c,\nonumber\\
    &&F(a0c010)=F(a1c010),\,\forall a,c,\nonumber\\
    &&F(abcxyz)\ge 0, \,\forall a,b,c,x,y,z.
\end{eqnarray}
The optimal test factor $F_{\mathsf{opt}}$ obtained from~\eqref{eq:TFoptCao} is presented in Table~\ref{tab:OptimalTF_Cao}. We introduce additional constraints to the optimisation routine in~\eqref{eq:TFoptCao}, specifically the TF ``locking" constraints $F(a0cxyz)=F(a1cxyz)$ for all values of $a,c$ and $xyz\equiv000,010$. This constraint results from our method of addressing the missing data problem. By enforcing these constraints we ensure that the test factors assign equal weight to the entries $F(a0cxyz)$ and $F(a1cxyz)$, which correspond to the entries $Q(a0cxyz)$ and $Q(a1cxyz)$ of the anticipated trial distribution for all values of $a$, $c$ and $xyz\equiv000,010$. This is because, as discussed in Appendix~\ref{a:MLNS_Cao}, in these cases only the sum $\sum_b Q(abcxyz)$ (and not the individual $Q(abcxyz)$ entries) can be considered to reflect the data of~\cite{Cao2022}, and so we ``lock" the test factors to not distinguish between the two $Q(abcxyz)$ for differing $b$.

\begin{figure}[!htb]
    \centering
    \includegraphics[width=1.0\linewidth]{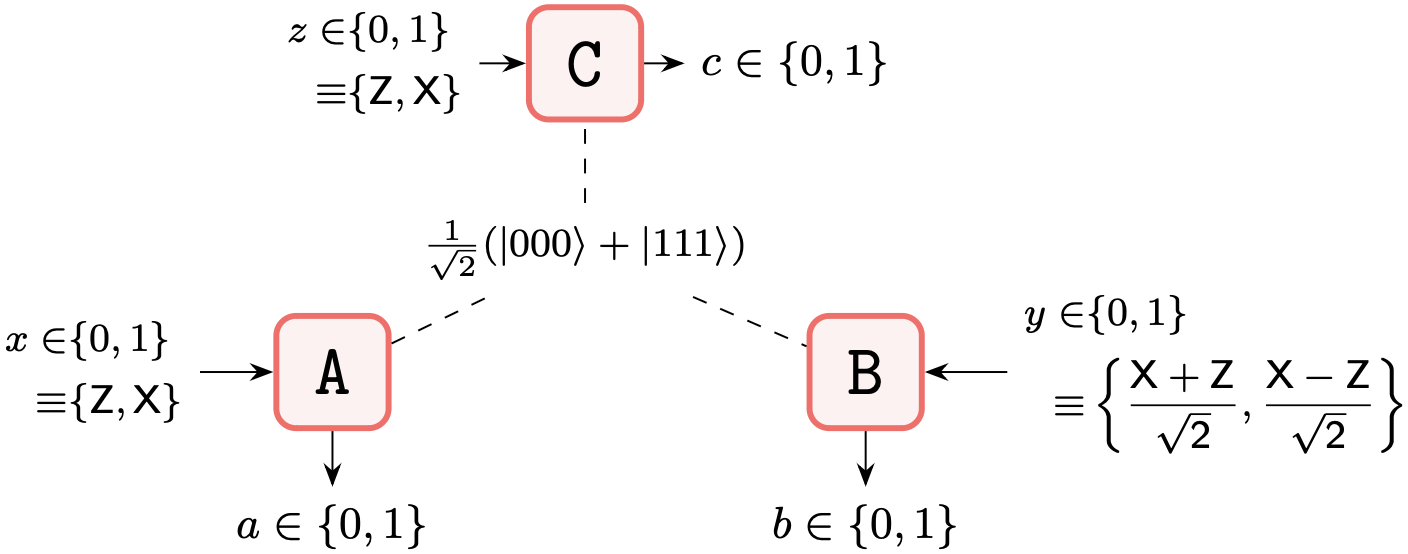}
    \caption{An ideal protocol resulting in a (perfect) GHZ behaviour that achieves the maximal violation $4+4\sqrt{2}$ for the linear witness in~\eqref{eq:Cao}. Parties $\mathsf{A,C}$ measure $\mathsf Z$ for $x,z=0$ and measure $\mathsf X$ for $x,z=1$, and party $\mathsf B$ measures $(\mathsf{X+Z})/\sqrt{2}$ for $y=0$ and $(\mathsf{X-Z})/\sqrt{2}$ for $y=1$. The choice of outcome labels $\{0,1\}$ is up to convention and can be used interchangeably with $\{+1,-1\}$ (which is conventionally the set of outcomes for Pauli measurements).}
    \label{fig:IdealProtocol}
\end{figure}

We now compare the strengths of evidence against $\mathcal{H}_{\mathsf{lo2sr}}$ for both experimental demonstrations in~\cite{Mao2022} and~\cite{Cao2022} with the strength of evidence for the distribution associated with a depolarised GHZ behaviour (assuming uniform settings distribution). The (perfect) GHZ behaviour maximally violating the witness in~\eqref{eq:Cao} results from the protocol shown in Figure~\ref{fig:IdealProtocol}. The protocol resulting in the GHZ behaviour that maximally violates the witness in~\eqref{eq:Mao} is same as that in Figure~\ref{fig:IdealProtocol} with the only difference that party $\mathsf B$ measures $(\mathsf{Z-X})/\sqrt{2}$ for the setting choice $y=1$.
While the perfect GHZ behaviours from the ideal protocols maximally violate the (respective) linear witnesses in~\eqref{eq:Mao} and~\eqref{eq:Cao}, thereby serving as a prototypical example of GTNL, practical considerations introduce noise, often modelled as a depolarising noise, as a result of which the pure state $\ket{\mathsf{GHZ}}=\frac{1}{\sqrt{2}}(\ket{000}+\ket{111})$ shared by the three parties is transformed into a mixed state. And so in the protocol displayed in Figure~\ref{fig:IdealProtocol} we replace the maximally entangled pure state $\ket{\mathsf{GHZ}}$ with a depolarised GHZ state which we model in terms of the visibility parameter $\theta\in[0,1]$ as $\Phi(\theta)=\theta\ket{\mathsf{GHZ}}\bra{\mathsf{GHZ}}+(1-\theta)I/8$ such that $\Phi(1)$ is the maximally entangled GHZ state and $\Phi(0)$ is the maximally mixed state $I/8$. The depolarised GHZ behaviour $\mathbf{P}_{\mathsf{GHZ},\theta}$ resulting from the protocol shown in Figure~\ref{fig:IdealProtocol} with $\ket{\mathsf{GHZ}}$ replaced with $\Phi(\theta)$ is as presented in Table~\ref{tab:noisyGHZ}. Similarly, we can obtain depolarised GHZ behaviour from the protocol (resulting in the perfect GHZ behaviour that maximally violates~\eqref{eq:Mao}) by replacing the state $\ket{\mathsf{GHZ}}$ with $\Phi(\theta)$.

\begin{figure}[!htb]
    \centering
    \includegraphics[width=1.0\linewidth]{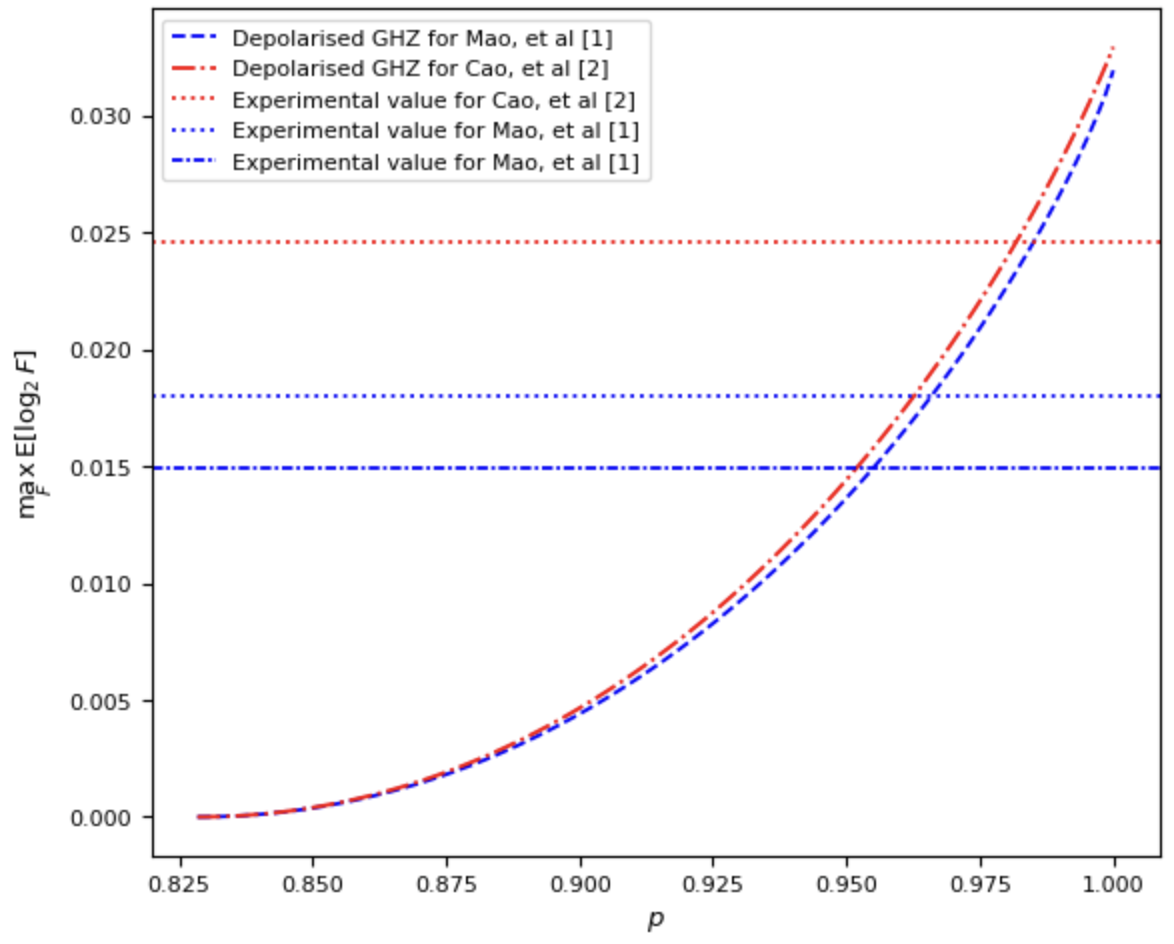}
    \caption{Strength of evidence against $\mathcal{H}_{\mathsf{lo2sr}}$ for the trial distribution $\mathbb{P}_{\mathsf{GHZ},\theta}$ associated with the depolarised GHZ behaviour $\mathbf{P}_{\mathsf{GHZ},\theta}$ (assuming a uniform settings distribution) with visibility $\theta$ varying in the interval $(2/(\sqrt{2}+1),1]$: The red curve displays strength of evidence for $\mathbb{P}_{\mathsf{GHZ},\theta}$ whose corresponding $\mathbf{P}_{\mathsf{GHZ},\theta}$ (as displayed in Table~\ref{tab:noisyGHZ}) violates the witness in~\eqref{eq:Cao} for $\theta>2/(\sqrt{2}+1)$. The blue curve displays strength of evidence for $\mathbb{P}_{\mathsf{GHZ},\theta}$ whose corresponding $\mathbf{P}_{\mathsf{GHZ},\theta}$ violates the witness in~\eqref{eq:Mao} for $\theta > 2/(\sqrt{2}+1)$. The horizontal lines denote the strength of evidence demonstrated by the experimental data from~\cite{Mao2022} (shown in blue) and~\cite{Cao2022} (shown in red). The blue lines correspond to values of $\approx0.014935$ and $\approx0.017984$ and the red line corresponds to a value of $\approx0.0245564$.}
    \label{fig:Max_logF_plot}
\end{figure}

\begin{table*}[!htb]
\caption{\label{tab:noisyGHZ} Depolarised GHZ behaviour resulting from the parties measuring a depolarised GHZ state $\Phi(\theta)=\theta\ket{\mathsf{GHZ}}\bra{\mathsf{GHZ}}+(1-\theta)I/8$ with visibility $\theta\in[0,1]$. Parties $\mathsf{A,C}$ measure $\mathsf{Z}$ and $\mathsf{X}$ for $x,z=0$ and $x,z=1$, respectively, and party $\mathsf{B}$ measures $(\mathsf{X+Z})/\sqrt{2}$ for $y=0$ and $(\mathsf{X-Z})/\sqrt{2}$ for $y=1$. This behaviour violates~\eqref{eq:Cao} for visibility $\theta>\frac{2}{\sqrt{2}+1}$, and for $\theta=1$ it is the perfect GHZ behaviour that achieves the maximum value of $4+4\sqrt{2}$ for the expression on the left hand side of~\eqref{eq:Cao}.}
\resizebox{0.95\textwidth}{!}{
\begin{ruledtabular}
\begin{tabular}{rrcccccccc}
\multicolumn{1}{r}{} & \multicolumn{1}{r}{} & \multicolumn{8}{c}{$abc$}\\
\cline{2-10}
\multicolumn{1}{r}{} & \multicolumn{1}{r}{} & \multicolumn{1}{c}{$000$} & \multicolumn{1}{c}{$001$} & \multicolumn{1}{c}{$010$} & \multicolumn{1}{c}{$011$} & \multicolumn{1}{c}{$100$} & \multicolumn{1}{c}{$101$} & \multicolumn{1}{c}{$110$} & \multicolumn{1}{c}{$111$} \\
\cline{2-10}
\multirow{8}{*}{\rotatebox[origin=c]{90}{$xyz$}} & $000$ & $\frac{1+\theta+\sqrt{2}\theta}{8}$ & $\frac{1-\theta}{8}$ & $\frac{1+\theta-\sqrt{2}\theta}{8}$ & $\frac{1-\theta}{8}$ & $\frac{1-\theta}{8}$ & $\frac{1+\theta-\sqrt{2}\theta}{8}$ & $\frac{1-\theta}{8}$ & 
  $\frac{1+\theta+\sqrt{2}\theta}{8}$ \\ 
  & $001$ & $\frac{2+\sqrt{2}\theta}{16}$ & $\frac{2+\sqrt{2}\theta}{16} $ & $\frac{2-\sqrt{2}\theta}{16} $ & $\frac{2-\sqrt{2}\theta}{16} $ & 
  $\frac{2-\sqrt{2}\theta}{16}$ & $\frac{2-\sqrt{2}\theta}{16} $ & $\frac{2+\sqrt{2}\theta}{16} $ & 
  $\frac{2+\sqrt{2}\theta}{16}$\\ 
  & $010$ & $\frac{1+\theta-\sqrt{2}\theta}{8} $ & $\frac{1-\theta}{8} $ & 
  $\frac{1+\theta+\sqrt{2}\theta}{8} $ & $\frac{1-\theta}{8} $ & $\frac{1-\theta}{8} $ & 
  $\frac{1+\theta+\sqrt{2}\theta}{8}$ & $\frac{1-\theta}{8}$ & 
  $\frac{1+\theta-\sqrt{2}\theta}{8}$ \\ 
  & $011$ & $\frac{2-\sqrt{2}\theta}{16}$ & $\frac{2-\sqrt{2}\theta}{16}$ & $\frac{2+\sqrt{2}\theta}{16}$ & $\frac{2+\sqrt{2}\theta}{16}$ & 
  $\frac{2+\sqrt{2}\theta}{16}$ & $\frac{2+\sqrt{2}\theta}{16}$ & $\frac{2-\sqrt{2}\theta}{16}$ & 
  $\frac{2-\sqrt{2}\theta}{16}$\\ 
  & $100$ & $\frac{2+\sqrt{2}\theta}{16}$ & $\frac{2-\sqrt{2}\theta}{16}$ & 
  $\frac{2-\sqrt{2}\theta}{16}$ & $\frac{2+\sqrt{2}\theta}{16}$ & $\frac{2+\sqrt{2}\theta}{16}$ & 
  $\frac{2-\sqrt{2}\theta}{16} $ & $\frac{2-\sqrt{2}\theta}{16} $ & 
  $\frac{2+\sqrt{2}\theta}{16}$ \\
  & $101$ & $\frac{2+\sqrt{2}\theta}{16}$ & $\frac{2-\sqrt{2}\theta}{16}$ & 
  $\frac{2-\sqrt{2}\theta}{16}$ & $\frac{2+\sqrt{2}\theta}{16}$ & $\frac{2-\sqrt{2}\theta}{16}$ & 
  $\frac{2+\sqrt{2}\theta}{16}$ & $\frac{2+\sqrt{2}\theta}{16}$ & 
  $\frac{2-\sqrt{2}\theta}{16}$ \\ 
  & $110$ & $\frac{2-\sqrt{2}\theta}{16}$ & $\frac{2+\sqrt{2}\theta}{16}$ & 
  $\frac{2+\sqrt{2}\theta}{16}$ & $\frac{2-\sqrt{2}\theta}{16}$ & $\frac{2-\sqrt{2}\theta}{16}$ & 
  $\frac{2+\sqrt{2}\theta}{16}$ & $\frac{2+\sqrt{2}\theta}{16}$ & 
  $\frac{2-\sqrt{2}\theta}{16}$ \\ 
  & $111$ & $\frac{2+\sqrt{2}\theta}{16}$ & $\frac{2-\sqrt{2}\theta}{16}$ & 
  $\frac{2-\sqrt{2}\theta}{16}$ & $\frac{2+\sqrt{2}\theta}{16}$ & $\frac{2-\sqrt{2}\theta}{16}$ & 
  $\frac{2+\sqrt{2}\theta}{16}$ & $\frac{2+\sqrt{2}\theta}{16}$ & $\frac{2-\sqrt{2}\theta}{16}$\\
\end{tabular}
\end{ruledtabular}}
\end{table*}
For visibility $\theta>2/(\sqrt{2}+1)$ the depolarised GHZ behaviour violates the witness in~\eqref{eq:Cao} (likewise for~\eqref{eq:Mao}). The red and blue curves in Figure~\ref{fig:Max_logF_plot} show a plot of the strength of evidence against $\mathcal{H}_{\mathsf{lo2sr}}$ for the trial distributions $\mathbb{P}_{\mathsf{GHZ},\theta}=\{P_{\mathsf{GHZ},\theta}(abc\lvert xyz)S(xyz)\}$, where $S(xyz)=1/8,\,\forall x,y,z$, and $\mathbf{P}_{\mathsf{GHZ},\theta}$ are the depolarised GHZ behaviours violating the respective inequalities in~\eqref{eq:Mao} and~\eqref{eq:Cao} for $\theta > 2/(\sqrt{2}+1)$.

The horizontal lines (in red and blue) depict the strength of evidence for the MLNS estimate of the empirical trial distribution obtained from the experimental data in~\cite{Mao2022} and~\cite{Cao2022}. The red (respectively, blue) curve in Figure~\ref{fig:Max_logF_plot} is obtained by performing the optimisation in~\eqref{eq:TFoptCao} (respectively, the one in~\eqref{eq:TFoptMao}) for $120$ equally spaced values of $\theta$ in the interval $(2/(\sqrt{2}+1),1]$ with $\mathsf{E}_{\mathbb{P}_{\mathsf{GHZ},\theta}}[\log_{2}F]$ as the objective quantity to be maximised. The plot shows a monotonic increase in the strength of evidence for both depolarised GHZ distributions $\mathbb{P}_{\mathsf{GHZ},\theta}$ as visibility $\theta$ increases beyond $2/(\sqrt{2}+1)$.

\section{Conclusion}\label{s:Conclusion}

In this paper, we have demonstrated a robust statistical method for analysing data obtained from experiments demonstrating genuine tripartite nonlocality. We proposed a computationally efficient modification to the test factor optimisation using an approximating polytope approach. It resulted in a significant reduction in the number of optimisation constraints while ensuring statistical validity. This methodology not only enhances the reliability of detecting genuine tripartite nonlocality in finite data from experiments but can also be applied to tests of properties involving violations of linear witnesses. 

Besides~\cite{Mao2022,Cao2022} another recent experimental demonstration of genuine tripartite nonlocality was reported in~\cite{JianWeiPan2022}. The tripartite scenario considered in this work is slightly different from the other two as party $\mathsf B$ has an extra setting, which makes it a $(2,2;2,3;2,2)$ Bell scenario. More generally, the $(2,|\mathcal{X}|;2,|\mathcal{Y}|;2,|\mathcal{Z}|)$ Bell scenario is a tripartite scenario with three parties $\mathsf{A,B,C}$ with settings choices from the sets $\mathcal{X,Y,Z}$, respectively, where each setting choice has two possible outcomes. We provide more details for the no-signalling set for this scenario in Appendix~\ref{a:NSreview}. The DI witness used in~\cite{JianWeiPan2022} is
\begin{equation}\label{eq:JWPan_ineq}
    \mathsf{CHSH}_{z=1}^{c=0} + \frac{4\mathsf{Same}-8}{1+\mathcal{M}_{\mathsf{C}}^{1}} \le 2,
\end{equation}
which is a combination of two Bell games: the CHSH game $\mathsf{CHSH}_{z=1}^{c=0}$ between parties $\mathsf A$ and $\mathsf B$ conditioned on $\mathsf C$ obtaining the outcomes $c=0$ for the measurement setting $z=1$, and the ``Same" game $\mathsf{Same}\coloneqq \mathcal{M}_{\mathsf{AB}}^{02}+\mathcal{M}_{\mathsf{BC}}^{20}$. We show in Appendix~\ref{a:JWPanDIWitness_Linear} that the witness in~\eqref{eq:JWPan_ineq} is linear in that it can be expressed in the form $\mathbf{B}\cdot\mathbf{P}\le\beta$ as shown in~\eqref{eq:JWPan_ineq_sumofprob}. Thus, our method could be used to analyse the data in a similar way for~\cite{JianWeiPan2022}, with the larger polytope defined by the no-signalling set and the witnessing inequality with the extra setting for party $\mathsf B$. However, since this scenario is experimentally more difficult to implement with no known offsetting benefit (the simpler inequalities of \eqref{eq:Mao} and \eqref{eq:Cao} appear to be more noise resistant \cite{Mao2022,Cao2022}), future experiments do not seem likely to test this inequality.

It is important to note that the experiments performed in~\cite{Mao2022,Cao2022} are subject to the locality loophole, and while the experiment in~\cite{JianWeiPan2022} ensured a space-like separation of the parties (thereby closing the locality loophole), all three experiments adopt a post-selection method in which they consider only those trials in which every party---each of which has two detectors---registers exactly one photon. Such experiments can only passively observe statistics consistent with a genuinely multipartite GHZ state in fundamentally limited fraction of trials---even if all components were improved to unit efficiency, there would always be multi-photon trials such that the non-post-selected statistics admit a model obeying the bipartite-only nonclassicality of Figure \ref{fig:3partytriangle}, or possibly even a fully classical model. (To illustrate, consider for example Figure 2a of \cite{JianWeiPan2022} where the experimental setup is indeed consistent with Figure \ref{fig:3partytriangle} as nonclassical resources comprise $A$-$C$ and $C$-$B$ bipartite quantum states; post-selection at $C$ allows retro-inference of GHZ behaviour in a fraction of trials but the full data could never rule out the Figure \ref{fig:3partytriangle} model.) While the PBR protocol is valid in the presence of memory effects and statistical fluctuations in finite data from experiments, the method presented here does not formalise or otherwise account for extra assumptions required by an experiment employing post-selection---the data is analysed as though produced by a non-post-selected experiment. Future experiments employing space-like separation and generating high-quality GHZ states without post-selection, such as through heralding---and employing sound statistical analysis such as the method presented here---will be able to take the next steps towards definitively exhibiting the phenomenon of genuine tripartite nonlocality.

\begin{acknowledgements}
    We thank Ya-Li Mao, Zheng-Da Li, Sixia Yu, and Jingyun Fan for sharing the count data from their experimental demonstration with us. This work was partially supported by NSF Grant No.~2328800 and AFOSR Grant No.~FA9550-20-1-0067.
\end{acknowledgements}

\nocite{*}

\bibliography{References}
\onecolumngrid
\appendix

\section{No-signalling set for the \texorpdfstring{$(2,|\mathcal{X}|;2,|\mathcal{Y}|;2,|\mathcal{Z}|)$}{Aa} Bell scenario}\label{a:NSreview}
We present a brief review of the no-signalling set of behaviours for a slightly more general three-party scenario. The scenario where parties have binary settings choice, i.e., $\mathcal{X,Y,Z}=\{0,1\}$ is then a special case. The $(2,|\mathcal{X}|;2,|\mathcal{Y}|;2,|\mathcal{Z}|)$ Bell scenario consists of three space-like separated parties $\mathsf{A},\mathsf{B}\text{ and }\mathsf{C}$ performing measurements with random and private input $x,y\text{ and }z$ chosen from their respective sets $\mathcal{X},\mathcal{Y}\text{ and }\mathcal{Z}$ and observing binary outcomes $a,b,c\in\{0,1\}$ for each measurement. The parties are bound by the no instantaneous-signalling principle, consistent with special relativity, which requires that $\mathsf{A}$ cannot signal to $\mathsf{B}$ or $\mathsf{C}$ (and cyclic permutations) together with the stronger condition that $\mathsf{A}$ cannot signal to the composite system $(\mathsf{B,C})$ (and cyclic permutations). It is expressed as the following equality constraints on the setting-conditional outcome probabilities $\mathbf{P}\coloneqq\{P(abc\lvert xyz)\}$:   
\begin{eqnarray}
    \sum_{a}P(abc\lvert xyz)&=&\sum_{a}P(abc\lvert x'yz),\,\forall b,c,x,x',y,z\label{eq:NS_overA}\\
    \sum_{b}P(abc\lvert xyz)&=&\sum_{b}P(abc\lvert xy'z),\,\forall a,c,x,y,y',z\label{eq:NS_overB}\\
    \sum_{c}P(abc\lvert xyz)&=&\sum_{c}P(abc\lvert xyz'),\,\forall a,b,x,y,z,z'.\label{eq:NS_overC}
\end{eqnarray}
Besides the above constraints, the behaviours satisfy the non-negativity and the normalisation constraints, expressed, respectively, as $P(abc\lvert xyz)\ge 0,\,\forall a,b,c,x,y,z$ and $\sum_{abc}P(abc\lvert xyz)=1,\,\forall x,y,z$.

Notice that we do not require a separate expression for the condition that the composite system $(\mathsf{A,B})$ cannot signal to $\mathsf{C}$ (and cyclic permutations) because it can be deduced from~\eqref{eq:NS_overA} and~\eqref{eq:NS_overB}, as demonstrated in~\cite{BarrettPironio2005}, which we show below for completeness.
\begin{eqnarray}\label{eq:NS_over_AB}
    \sum_{a,b}P(abc\lvert xyz)&=&\sum_{a,b}P(abc\lvert x'yz),\,\forall c,x,x',y,z\nonumber \\
    &=&\sum_{a,b}P(abc\lvert x'y'z),\,\forall c,x,x',y,y',z.
\end{eqnarray}
The $8\abs{\mathcal{X}}\abs{\mathcal{Y}}\abs{\mathcal{Z}}$ inequalities given by the non-negativity constraints describe closed half-spaces. Equalities given by the $\abs{\mathcal{X}}\abs{\mathcal{Y}}\abs{\mathcal{Z}}$ normalisation constraints and the $4\left[(\abs{\mathcal{X}}-1)\abs{\mathcal{Y}}\abs{\mathcal{Z}}+\abs{\mathcal{X}}(\abs{\mathcal{Y}}-1)\abs{\mathcal{Z}} + \abs{\mathcal{X}}\abs{\mathcal{Y}}(\abs{\mathcal{Z}}-1)\right]$ no-signalling constraints describe hyperplanes. The intersection of the closed half-spaces and the hyperplanes gives the no-signalling polytope. The ambient space containing $\Xi_{\mathsf{ns}}$ is $\mathbb{R}^{t}$, where $t=8\abs{\mathcal{X}}\abs{\mathcal{Y}}\abs{\mathcal{Z}}$; however, it is known that after subtracting redundancies, $\Xi_{\mathsf{ns}}$ is contained in a lower-dimensional affine space, the dimension of which is $\mathrm{dim}(\Xi_{\mathsf{ns}})=(\abs{\mathcal{X}}+1)(\abs{\mathcal{Y}}+1)(\abs{\mathcal{Z}}+1)-1$ (see Theorem 3.1 in~\cite{PironioPhDThesis}). For the scenario where each party has binary input choices, i.e., $\mathcal{X},\mathcal{Y},\mathcal{Z}=\{0,1\}$, the no-signalling set $\Xi_{\mathsf{ns}}$ lies in a $26$-dimensional affine subspace of $\mathbb{R}^{64}$ and consists of $53,856$ extreme points, of which $64$ are local deterministic and the remaining grouped under $45$ equivalence classes (see~\cite{ExtrCorrTripartite_Pironio} for a complete list of the representative behaviours for each equivalence class along with the number of distinct behaviours obtained by a relabelling of outcomes, settings and parties). The expressions as defined in~\eqref{eq:MarginCorrA}--\eqref{eq:FullCorr} for the marginalised and the full correlators can equivalently be written as shown below:
\begin{equation}\label{eq:CorrVector}
    \begin{bmatrix}
     1 \\ \mathcal{M}_{\mathsf{A}}^{x} \\ \mathcal{M}_{\mathsf{B}}^{y} \\ \mathcal{M}_{\mathsf{C}}^{z} \\ \mathcal{M}_{\mathsf{AB}}^{xy} \\ \mathcal{M}_{\mathsf{BC}}^{yz} \\ \mathcal{M}_{\mathsf{AC}}^{xz} \\ \mathcal{C}_{\mathsf{ABC}}^{xyz}
 \end{bmatrix} = \begin{bmatrix}
 1 & 1 & 1 & 1 & 1 & 1 & 1 & 1 \\
 1 & 1 & 1 & 1 & -1 & -1 & -1 & -1 \\
 1 & 1 & -1 & -1 & 1 & 1 & -1 & -1 \\
 1 & -1 & 1 & -1 & 1 & -1 & 1 & -1 \\
 1 & 1 & -1 & -1 & -1 & -1 & 1 & 1 \\
 1 & -1 & -1 & 1 & 1 & -1 & -1 & 1 \\
 1 & -1 & 1 & -1 & -1 & 1 & -1 & 1 \\
 1 & -1 & -1 & 1 & -1 & 1 & 1 & -1 \end{bmatrix}\begin{bmatrix}
    P(000\lvert xyz) \\ P(001\lvert xyz) \\ P(010\lvert xyz) \\ P(011\lvert xyz) \\ P(100\lvert xyz) \\ P(101\lvert xyz) \\ P(110\lvert xyz) \\ P(111\lvert xyz) \\
    \end{bmatrix}.
\end{equation}    
The following compact expression for behaviour $\{P(abc\lvert xyz)\}$, as a parametrisation of the outcomes and correlators, is obtained from the inverse of~\eqref{eq:CorrVector}.
\begin{eqnarray}\label{eq:Behaviour_expr1}
    P(abc\lvert xyz) &=& \frac{1}{8}\Big[1 + (-1)^{a}\mathcal{M}_{\mathsf{A}}^{x} + (-1)^{b}\mathcal{M}_{\mathsf{B}}^{y}+ (-1)^{c}\mathcal{M}_{\mathsf{C}}^{z} + (-1)^{a\oplus b}\mathcal{M}_{\mathsf{AB}}^{xy} +  (-1)^{b\oplus c}\mathcal{M}_{\mathsf{BC}}^{yz} + \\ \nonumber
    && (-1)^{a\oplus c}\mathcal{M}_{\mathsf{AC}}^{xz} + (-1)^{a\oplus b\oplus c}\mathcal{C}_{\mathsf{ABC}}^{xyz}\Big].
\end{eqnarray}

\section{Maximum-likelihood no-signalling (MLNS) estimate}\label{a:MLNSestimate}
In this section we present the relevant details on obtaining a maximum-likelihood no-signalling estimate of an empirical trial distribution which is derived from the raw experimental data as $f_{\mathsf{emp}}(abcxyz)=\frac{N(abc\lvert xyz)}{\sum_{abc}N(abc\lvert xyz)}S(xyz)$, where $N(abc\lvert xyz)$ represents the count of outcome combinations $a,b,c$ given the measurement settings combinations $x,y,z$. Owing to statistical fluctuations, $\{f_{\mathsf{emp}}(abcxyz)\}$ may not adhere to the no-signalling and measurement settings constraints, potentially exhibiting weak signalling effects. So we obtain an estimate of the trial distribution that satisfies these constraints. One method to obtain such an estimate is based on maximum-likelihood as mentioned in~\cite{ZhangGlancyKnill2011} (refer to the paragraph immediately following (A2) in Section 2 of the appendix). Assuming $n$ iid trials, the likelihood of the trial data with empirical frequencies $\{f_{\mathsf{emp}}(abcxyz)\}$, given that the true distribution is $\mathbb{Q}\coloneqq\{Q(abcxyz)\}$, is proportional to $\prod_{abcxyz}Q(abcxyz)^{nf_{\mathsf{emp}}(abcxyz)}$. The base-2 logarithm of this (ignoring the constant factor $n$) is the objective quantity to be maximised as shown below:
\begin{eqnarray}\label{eq:MLNSestimate}
    \max_{\mathbb Q}\,\, &&\sum_{abcxyz}f_{\mathsf{emp}}(abcxyz)\log_{2}Q(abcxyz)\nonumber \\
    \text{subject to}\,\, &&\sum_{a}Q(abc\lvert xyz)=\sum_{a}Q(abc\lvert x'yz),\,\forall b,c,x,x',y,z\nonumber \\
                      &&\sum_{b}Q(abc\lvert xyz)=\sum_{b}Q(abc\lvert xy'z),\,\forall a,c,x,y,y',z\nonumber \\
                      &&\sum_{c}Q(abc\lvert xyz)=\sum_{c}Q(abc\lvert xyz'),\,\forall a,b,x,y,z,z'\nonumber \\
                      &&\sum_{a,b,c}Q(abcxyz)=S(xyz),\,\forall x,y,z.
\end{eqnarray}
In~\eqref{eq:MLNSestimate}, the first three constraints represent the absence of signalling, followed by the requirement that $\mathbb Q$ must conform to the settings distribution. The result obtained from~\eqref{eq:MLNSestimate} is the maximum-likelihood no-signalling (MLNS) estimate $\hat{\mathbb Q}$ of the empirical trial distribution. Given the concavity of the objective function, the maximisation routine in~\eqref{eq:MLNSestimate} can be implemented using a standard convex optimisation programme. It must be noted, however, that when implementing~\eqref{eq:MLNSestimate} additional constraints pertaining to non-negativity must be included since the estimate $\hat{\mathbb Q}$ that we are seeking is a distribution of the experiment results. The optimisation routine to obtain useful TFs is then performed with respect to $\hat{\mathbb Q}$. 

\subsection{MLNS estimate for experimental data from \texorpdfstring{~\cite{Mao2022}}{M}}\label{a:MLNS_Mao}

For the experimental demonstration in~\cite{Mao2022} we obtained two different count datasets from the authors of that work \cite{PrivateComm}. We first consider the count dataset for a four-party GHZ experiment implementing a protocol involving the state $\ket{\Phi}=\frac{1}{\sqrt{2}}(\ket{0000}+\ket{1111})$. The four-party data must be processed to obtain the three-party data. The data for a three-photon GHZ experiment is derived from a four-photon GHZ experiment by measuring the third photon in the $\mathsf{X}$ basis and conditioning on the ``$+$" outcome~\cite{PrivateComm,Mao2022}. And so in Table~\ref{tab:Rawdata_Mao_1} we present only that portion of the four-party data from~\cite{PrivateComm} where the third party's measurement setting is $\mathsf{X}$. Columns corresponding to outcome combinations where the third entry is a ``$-$" are then ignored.

\begin{table*}[!htb]
\caption{\label{tab:Rawdata_Mao_1} Count data obtained from~\cite{PrivateComm} for the four-party GHZ experiment performed in~\cite{Mao2022}. The columns labelled by the various combinations of $\{+,-\}$ denote the measurement outcomes and the rows denote the (Pauli) measurements. The empirical trial distribution is obtained from the three-party data processed from this four-party data, by ignoring the shaded columns where the third party observes ``$-$"}.
\footnotesize
\resizebox{\columnwidth}{!}{%
\begin{ruledtabular}
\begin{tabular}{cccaaccaaccaaccaa}
\multicolumn{1}{c}{} & \multicolumn{1}{c}{$\scaleto{++++}{4pt}$} & \multicolumn{1}{c}{$\scaleto{+++-}{4pt}$} & \multicolumn{1}{c}{$\scaleto{++-+}{4pt}$} & \multicolumn{1}{c}{$\scaleto{++--}{4pt}$} & \multicolumn{1}{c}{$\scaleto{+-++}{4pt}$} & \multicolumn{1}{c}{$\scaleto{+-+-}{4pt}$} & \multicolumn{1}{c}{$\scaleto{+--+}{4pt}$} & \multicolumn{1}{c}{$\scaleto{+---}{4pt}$} &
\multicolumn{1}{c}{$\scaleto{-+++}{4pt}$} & \multicolumn{1}{c}{$\scaleto{-++-}{4pt}$} & \multicolumn{1}{c}{$\scaleto{-+-+}{4pt}$} & \multicolumn{1}{c}{$\scaleto{-+--}{4pt}$} & \multicolumn{1}{c}{$\scaleto{--++}{4pt}$} & \multicolumn{1}{c}{$\scaleto{--+-}{4pt}$} & \multicolumn{1}{c}{$\scaleto{---+}{4pt}$} & \multicolumn{1}{c}{$\scaleto{----}{4pt}$} \\
    \cline{2-17} $\scaleto{\mathsf{Z}\frac{\mathsf{X+Z}}{\sqrt 2}\mathsf{XZ}}{8pt}$ & 550 & 3 & 541 & 2 & 114 & 3 & 130 & 1 & 4 & 113 & 4 & 113 & 5 & 826 & 3 & 923 \\ $\scaleto{\mathsf{Z}\frac{\mathsf{X+Z}}{\sqrt 2}\mathsf{XX}}{8pt}$ & 257 & 282 & 306 & 312 & 64 & 67 & 69 & 78 & 51 & 51 & 52 & 67 & 485 & 466 & 460 & 426 \\  $\scaleto{\mathsf{Z}\frac{\mathsf{X-Z}}{\sqrt 2}\mathsf{XZ}}{8pt}$ & 87 & 2 & 82 & 0 & 671 & 1 & 737 & 4 & 2 & 639 & 0 & 566 & 4 & 179 & 2 & 158
 \\ $\scaleto{\mathsf{Z}\frac{\mathsf{X-Z}}{\sqrt 2}\mathsf{XX}}{8pt}$ & 45 & 64 & 55 & 61 & 373 & 365 & 391 & 355 & 309 & 322 & 290 & 295 & 81 & 73 & 72 & 91
 \\ $\scaleto{\mathsf{X}\frac{\mathsf{X+Z}}{\sqrt 2}\mathsf{XZ}}{8pt}$ & 259 & 55 & 310 & 53 & 70 & 394 & 69 & 409 & 285 & 50 & 309 & 51 & 69 & 392 & 61 & 408
 \\ $\scaleto{\mathsf{X}\frac{\mathsf{X+Z}}{\sqrt 2}\mathsf{XX}}{8pt}$ & 275 & 63 & 57 & 302 & 104 & 401 & 393 & 86 & 54 & 293 & 252 & 57 & 434 & 94 & 74 & 411
 \\ $\scaleto{\mathsf{X}\frac{\mathsf{X-Z}}{\sqrt 2}\mathsf{XZ}}{8pt}$ & 44 & 280 & 68 & 316 & 377 & 80 & 363 & 68 & 48 & 304 & 84 & 328 & 370 & 75 & 424 & 83
 \\ $\scaleto{\mathsf{X}\frac{\mathsf{X-Z}}{\sqrt 2}\mathsf{XX}}{8pt}$ & 280 & 58 & 60 & 298 & 68 & 373 & 363 & 72 & 73 & 329 & 302 & 67 & 354 & 59 & 103 & 396
 \\
\end{tabular}
\end{ruledtabular}}
\end{table*}
After this step, there is one more manipulation needed. In the original three-party protocol described in ~\cite{Mao2022}, party $\mathsf{B}$ performs the measurements $\frac{\mathsf{Z+X}}{\sqrt{2}}$ and $\frac{\mathsf{Z-X}}{\sqrt{2}}$. However, the four-party data we have from ~\cite{PrivateComm} uses the measurements $\frac{\mathsf{X+Z}}{\sqrt{2}}$ and $\frac{\mathsf{X-Z}}{\sqrt{2}}$ for the second photon. So for a given outcome combination of the first, third, and fourth photons, we swap the count entry where the outcome for the second photon is ``$+$" with the count entry where the outcome for the second photon is ``$-$". Table~\ref{tab:EmpiricalDist_Mao_1} shows the empirical trial distribution obtained from the three-party count data derived from the four-party data. In obtaining the empirical trial distribution, we have assumed a uniform measurement settings distribution $S(xyz)=1/8,\,\forall x,y,z$.

\begin{table*}[!htb]
\caption{\label{tab:EmpiricalDist_Mao_1} Empirical trial distribution $f_{\mathsf{emp}}(abcxyz)$ for the raw experimental data in Table~\ref{tab:Rawdata_Mao_1} assuming a uniform settings distribution $S(xyz)=1/8,\,\forall x,y,z$.}
\resizebox{0.95\textwidth}{!}{%
\begin{ruledtabular}
\begin{tabular}{rrcccccccc}
\multicolumn{1}{r}{} & \multicolumn{1}{r}{} & \multicolumn{8}{c}{$abc$}\\
\cline{2-10}
\multicolumn{1}{r}{} & \multicolumn{1}{r}{} & \multicolumn{1}{c}{$000$} & \multicolumn{1}{c}{$001$} & \multicolumn{1}{c}{$010$} & \multicolumn{1}{c}{$011$} & \multicolumn{1}{c}{$100$} & \multicolumn{1}{c}{$101$} & \multicolumn{1}{c}{$110$} & \multicolumn{1}{c}{$111$} \\
\cline{2-10}
\multirow{8}{*}{\rotatebox[origin=c]{90}{$xyz$}} & $000$ & 0.04249073 & 0.00023177 & 0.00880717 & 0.00023177 & 0.00030902 & 0.00872991 &  0.00038628 & 0.06381335 \\
 & $001$ & 0.01864481 & 0.0204585 & 0.00464306 & 0.00486071 & 0.00369994 & 0.00369994 & 0.03518572 & 0.03380731 \\
 & $010$ & 0.05291798 & 0.00007886 & 0.0068612 &  0.00015773 & 0.00031546 & 0.01411672 & 0.00015773 & 0.05039432 \\
 & $011$ & 0.02856924 & 0.0279565 & 0.00344669 & 0.00490196 & 0.00620404 & 0.0055913 & 0.02366728 & 0.02466299 \\
 & $100$ & 0.02056861 & 0.00436785 & 0.00555909 & 0.03128971 & 0.02263342 & 0.00397078 & 0.00547967 & 0.03113088 \\
 & $101$ & 0.02000873 & 0.00458382 & 0.00756694 & 0.02917637 & 0.00392899 & 0.02131839 & 0.03157742 & 0.00683935 \\
 & $110$ & 0.02986375 & 0.00633714 & 0.00348542 & 0.02217997 & 0.02930925 & 0.00594106 & 0.00380228 & 0.02408112 \\
 & $111$ & 0.0053325 & 0.02925031 & 0.02195734 & 0.00454831 & 0.02776035 & 0.00462673 & 0.00572459 & 0.02579987 \\
\end{tabular}
\end{ruledtabular}}
\end{table*}
For the empirical trial distribution in Table~\ref{tab:EmpiricalDist_Mao_1} we use the maximisation routine in~\eqref{eq:MLNSestimate} to find the MLNS estimate $\hat{\mathbb Q}$. The optimisation routine in~\eqref{eq:TFoptMao} is then performed with respect to the distribution in Table~\ref{tab:MLNSest_Mao_1} from which we obtain the optimal test factor given in Table~\ref{tab:OptimalTF_Mao}.

\begin{table*}[!htb]
\caption{\label{tab:MLNSest_Mao_1} MLNS estimate for the empirical distribution in Table~\ref{tab:EmpiricalDist_Mao_1} from the experiment in~\eqref{eq:Mao} conforming to the uniform settings distribution $S(xyz)=1/8,\,\forall x,y,z$.}
\resizebox{0.95\textwidth}{!}{%
\begin{ruledtabular}
\begin{tabular}{rrcccccccc}
\multicolumn{1}{r}{} & \multicolumn{1}{r}{} & \multicolumn{8}{c}{$abc$}\\
\cline{2-10}
\multicolumn{1}{r}{} & \multicolumn{1}{r}{} & \multicolumn{1}{c}{$000$} & \multicolumn{1}{c}{$001$} & \multicolumn{1}{c}{$010$} & \multicolumn{1}{c}{$011$} & \multicolumn{1}{c}{$100$} & \multicolumn{1}{c}{$101$} & \multicolumn{1}{c}{$110$} & \multicolumn{1}{c}{$111$} \\
\cline{2-10}
\multirow{8}{*}{\rotatebox[origin=c]{90}{$xyz$}} & $000$ & 0.04600651 & 0.000163 & 0.01110175 & 0.00015666 & 0.00018501 & 0.00717552 & 0.00038714 & 0.05982447 \\
 & $001$ & 0.02213168 & 0.02403783 & 0.00572623 & 0.00553217 & 0.00357774 & 0.00378272 & 0.03074134 & 0.02947028 \\
 & $010$ & 0.05061083 & 0.00008486 & 0.00649748 & 0.00023468 & 0.00037697 & 0.0141965 & 0.00019509 & 0.05280355 \\
 & $011$ & 0.02510992 & 0.02558579 & 0.00274798 & 0.0039842 & 0.00775529 & 0.00681817 & 0.0265638 & 0.02643483 \\
 & $100$ & 0.02256865 & 0.00375386 & 0.00593747 & 0.02931907 & 0.02362281 & 0.00358466 & 0.00555143 & 0.03066206 \\
 & $101$ & 0.02153297 & 0.00478953 & 0.00709426 & 0.02816228 & 0.00417644 & 0.02303102 & 0.02937332 & 0.00684017 \\
 & $110$ & 0.0254654 & 0.00769418 & 0.00304072 & 0.02537874 & 0.0255224 & 0.0065872 & 0.00365184 & 0.02765951 \\
 & $111$ & 0.00514717 & 0.02801242 & 0.02348007 & 0.00493939 & 0.02771805 & 0.00439155 & 0.00583171 & 0.02547964 \\
\end{tabular}
\end{ruledtabular}}
\end{table*}

In~\cite{ZhangGlancyKnill2011}, the authors address an issue of zero counts in raw data when testing local realism using the full-PBR protocol. In the presence of zero-valued counts in the raw count data, the empirical trial distribution obtained from it retains some zero-valued entries, in which case the PBR method may assign the value zero to the test factor score of the corresponding settings-outcome combinations. Then if the corresponding settings-outcome combinations occur in a subsequent trial, the $p$-value is set to $1$ without any possibility for a later recovery. These zero counts result in zero frequencies for certain combinations of measurement settings and outcomes in the empirical distribution, adversely affecting the calculation of full-PBR $p$-values. We encounter this issue of zero counts in the second dataset obtained from~\cite{PrivateComm}. The count data presented in Table~\ref{tab:Rawdata_Mao_2} is from the experiment in~\cite{Mao2022} implementing a protocol involving the state $\ket{\Phi}=\sqrt{\frac{1.5}{2}}\ket{0000}+\sqrt{\frac{0.5}{2}}\ket{1111}$, which is an unbalanced $4$-qubit GHZ state. The three-party data derived from the four-party data retains some occurrences of zero counts; specifically, the entries for the settings-outcomes combinations $\left(+++-,\mathsf{Z\frac{X+Z}{\sqrt 2}XZ}\right)$, $\left(+++-,\mathsf{Z\frac{X-Z}{\sqrt 2}XZ}\right)$ and $\left(-+++,\mathsf{Z\frac{X+Z}{\sqrt 2}XZ}\right)$. (The data processing steps to derive the three-party data is same as the previous dataset.) We then follow the approach outlined in~\cite{ZhangGlancyKnill2011} (refer to section 2 of the appendix) to deal with the issue of having zero frequencies in the empirical distribution obtained from the three-party data: we first use the optimisation routine in~\eqref{eq:MLNSestimate} to obtain a first no-signalling estimate $\hat{\mathbb{Q}}_{0}$ for the true distribution $\mathbb{Q}$, with which we mix in a distribution that has no zero probabilities with a weight approaching zero as $n$ grows. That is, the adjusted estimate $\hat{\mathbb{Q}}_{1}$ is given as follows:
\begin{equation}\label{eq:AdjustedMLNSestimate}
    \hat{Q}_{1}(abcxyz) = \frac{n}{n+1}\hat{Q}_{0}(abcxyz) + \frac{1}{n+1}\frac{1}{8}S(xyz).
\end{equation}
In~\eqref{eq:AdjustedMLNSestimate}, we have mixed the first estimate $\hat{\mathbb Q}_{0}$ with a distribution that, conditionally on the settings, is uniform and conforms to the settings distribution $S(xyz)$. In our implementations, we assume a uniform settings distribution, i.e., we substitute $S(xyz)$ with $1/8$ for all $x,y,z$. The empirical distribution obtained from the three-party count data, which is derived from the four-party data shown in Table~\ref{tab:Rawdata_Mao_2}, is presented in Table~\ref{tab:EmpiricalDist_Mao_2}. The adjusted MLNS estimate, according to~\eqref{eq:AdjustedMLNSestimate}, is presented in Table~\ref{tab:MLNSest_Mao_2}. 

\begin{table*}[!htb]
\caption{\label{tab:Rawdata_Mao_2}Count data for the four-party GHZ experiment in~\cite{Mao2022} involving an unbalanced GHZ state. The columns and rows, respectively, denote the outcomes and measurements for the parties. To obtain the three-party data we ignore the columns in grey. The remaining data is used as is with  the measurements $(\mathsf{Z},\mathsf{X})$ corresponding to the settings choices $(0,1)$ for parties $\mathsf{A}$ and $\mathsf{C}$ and the measurements $\left(\frac{\mathsf{Z}+\mathsf{X}}{\sqrt{2}},\frac{\mathsf{Z}-\mathsf{X}}{\sqrt{2}}\right)$ corresponding to the settings choices $(0,1)$ for party $\mathsf{B}$.}
\footnotesize
\resizebox{\columnwidth}{!}{%
\begin{ruledtabular}
\begin{tabular}{cccaaccaaccaaccaa}
\multicolumn{1}{c}{} & \multicolumn{1}{c}{$\scaleto{++++}{4pt}$} & \multicolumn{1}{c}{$\scaleto{+++-}{4pt}$} & \multicolumn{1}{c}{$\scaleto{++-+}{4pt}$} & \multicolumn{1}{c}{$\scaleto{++--}{4pt}$} & \multicolumn{1}{c}{$\scaleto{+-++}{4pt}$} & \multicolumn{1}{c}{$\scaleto{+-+-}{4pt}$} & \multicolumn{1}{c}{$\scaleto{+--+}{4pt}$} & \multicolumn{1}{c}{$\scaleto{+---}{4pt}$} &
\multicolumn{1}{c}{$\scaleto{-+++}{4pt}$} & \multicolumn{1}{c}{$\scaleto{-++-}{4pt}$} & \multicolumn{1}{c}{$\scaleto{-+-+}{4pt}$} & \multicolumn{1}{c}{$\scaleto{-+--}{4pt}$} & \multicolumn{1}{c}{$\scaleto{--++}{4pt}$} & \multicolumn{1}{c}{$\scaleto{--+-}{4pt}$} & \multicolumn{1}{c}{$\scaleto{---+}{4pt}$} & \multicolumn{1}{c}{$\scaleto{----}{4pt}$} \\
\cline{2-17} $\scaleto{\mathsf{Z}\frac{\mathsf{X+Z}}{\sqrt{2}}\mathsf{X}\mathsf{Z}}{8pt}$ & 204 & 0	& 282 & 1 & 73 & 2 & 69 & 0 & 0 & 24 & 2 & 18 & 2 & 138 & 0 & 155 \\ $\scaleto{\mathsf{Z}\frac{\mathsf{X+Z}}{\sqrt{2}}\mathsf{X}\mathsf{X}}{8pt}$ & 110 & 111 & 136 & 133 & 27 & 37 & 50 & 31 & 10 & 11 & 7 & 5 & 85 & 75 & 70 & 69 \\  $\scaleto{\mathsf{Z}\frac{\mathsf{X-Z}}{\sqrt{2}}\mathsf{X}\mathsf{Z}}{8pt}$ & 36 & 0 & 59 & 0 & 289 & 2 & 374 & 3 & 1 & 121 & 0 & 118 & 2 & 34 & 0 & 25 \\ $\scaleto{\mathsf{Z}\frac{\mathsf{X-Z}}{\sqrt{2}}\mathsf{X}\mathsf{X}}{8pt}$ & 23 & 20 & 19 & 30 & 178 & 189 & 198 & 230 & 50 & 70 & 70 & 47 & 18 & 14 & 11 & 17 \\ $\scaleto{\mathsf{X}\frac{\mathsf{X+Z}}{\sqrt{2}}\mathsf{X}\mathsf{Z}}{8pt}$ & 129 & 6 & 128 & 5 & 30 & 83 & 36 & 62 & 109 & 10 & 130 & 10 & 31 & 79 & 47 & 62 \\ $\scaleto{\mathsf{X}\frac{\mathsf{X+Z}}{\sqrt{2}}\mathsf{X}\mathsf{X}}{8pt}$ & 125 & 51 & 36 & 126 & 7 & 112 & 101 & 7 & 56 & 119 & 133 & 39 & 111 & 11 & 6 & 137 \\ $\scaleto{\mathsf{X}\frac{\mathsf{X-Z}}{\sqrt{2}}\mathsf{X}\mathsf{Z}}{8pt}$ & 12 & 61 & 21 & 55 & 150 & 11 & 183 & 11 & 22 & 50 & 19 & 48 & 171 & 7 & 205 & 11 \\ $\scaleto{\mathsf{X}\frac{\mathsf{X-Z}}{\sqrt{2}}\mathsf{X}\mathsf{X}}{8pt}$ & 81 & 6 & 8 & 67 & 62 & 141 & 145 & 66 & 6 & 75 & 67 & 7 & 156 & 53 & 66 & 172 \\
\end{tabular}
\end{ruledtabular}}
\end{table*}

\begin{table*}[!htb]
\caption{\label{tab:EmpiricalDist_Mao_2}Empirical trial distribution for the raw experimental data in Table~\ref{tab:Rawdata_Mao_2} assuming a uniform settings distribution $S(xyz)=1/8,\,\forall x,y,z$.}
\resizebox{0.95\textwidth}{!}{%
\begin{ruledtabular}
\begin{tabular}{rrcccccccc}
\multicolumn{1}{r}{} & \multicolumn{1}{r}{} & \multicolumn{8}{c}{$abc$}\\
\cline{2-10}
\multicolumn{1}{r}{} & \multicolumn{1}{r}{} & \multicolumn{1}{c}{$000$} & \multicolumn{1}{c}{$001$} & \multicolumn{1}{c}{$010$} & \multicolumn{1}{c}{$011$} & \multicolumn{1}{c}{$100$} & \multicolumn{1}{c}{$101$} & \multicolumn{1}{c}{$110$} & \multicolumn{1}{c}{$111$} \\
\cline{2-10}
\multirow{8}{*}{\rotatebox[origin=c]{90}{$xyz$}} & $000$ & 0.05756208 & 0. & 0.02059819 & 0.00056433 & 0. & 0.00677201 & 0.00056433 & 0.03893905 \\
 & $001$ & 0.02950644 & 0.02977468 & 0.00724249 & 0.00992489 & 0.0026824 &  0.00295064 & 0.02280043 & 0.02011803 \\
 & $010$ & 0.07448454 & 0.00051546 & 0.00927835 & 0. & 0.00051546 & 0.00876289 & 0.00025773 & 0.03118557 \\
 & $011$ & 0.03959075 & 0.04203737 & 0.00511566 & 0.0044484 & 0.00400356 & 0.00311388 & 0.011121 & 0.0155694 \\
 & $100$ & 0.03380503 & 0.00157233 & 0.00786164 & 0.02175052 & 0.02856394 & 0.00262055 & 0.00812369 & 0.02070231\\
 & $101$ & 0.02639358 & 0.01076858 & 0.00147804 & 0.02364865 & 0.01182432 & 0.02512669 & 0.0234375 & 0.00232264 \\
 & $110$ & 0.03873967 & 0.00284091 & 0.00309917 & 0.01575413 & 0.04416322 & 0.00180785 & 0.00568182 & 0.01291322 \\
 & $111$ & 0.01336207 & 0.03038793 & 0.0174569 & 0.0012931 & 0.03362069 & 0.01142241 & 0.0012931 & 0.01616379 \\
\end{tabular}
\end{ruledtabular}}
\end{table*}

\begin{table*}[!htb]
\caption{\label{tab:MLNSest_Mao_2} Adjusted MLNS estimate for the empirical trial distribution in Table~\ref{tab:EmpiricalDist_Mao_2} conforming to a uniform settings distribution. The estimate is obtained according to~\eqref{eq:AdjustedMLNSestimate}, where $\hat{\mathbb{Q}}_{0}$ is a first estimate obtained using~\eqref{eq:MLNSestimate} with respect to Table~\ref{tab:EmpiricalDist_Mao_2}.}
\resizebox{0.95\textwidth}{!}{%
\begin{ruledtabular}
\begin{tabular}{rrcccccccc}
\multicolumn{1}{r}{} & \multicolumn{1}{r}{} & \multicolumn{8}{c}{$abc$}\\
\cline{2-10}
\multicolumn{1}{r}{} & \multicolumn{1}{r}{} & \multicolumn{1}{c}{$000$} & \multicolumn{1}{c}{$001$} & \multicolumn{1}{c}{$010$} & \multicolumn{1}{c}{$011$} & \multicolumn{1}{c}{$100$} & \multicolumn{1}{c}{$101$} & \multicolumn{1}{c}{$110$} & \multicolumn{1}{c}{$111$} \\
\cline{2-10}
\multirow{8}{*}{\rotatebox[origin=c]{90}{$xyz$}} & $000$ & 0.06625581 & 0.00000023 & 0.01811245 & 0.00047703 & 0.00001538 & 0.00495332 & 0.00062414 & 0.03456163\\
 & $001$ & 0.03419785 & 0.03205819 & 0.00805397 & 0.01053551 & 0.00227225 & 0.00269645
 & 0.01790048 & 0.01728529\\
 & $010$ & 0.07628384 & 0.00046728 & 0.00808442 & 0.00000996 & 0.00046337 & 0.00806587
 & 0.00017618 & 0.03144907\\
 & $011$ & 0.0378644 & 0.03888672 & 0.00438742 & 0.00370697 & 0.00525107 & 0.00327817 & 0.01492167 & 0.01670358\\
 & $100$ & 0.03348011 & 0.00208906 & 0.00817827 & 0.01873282 & 0.03279109 & 0.00286447
 & 0.01055833 & 0.01630584\\
 & $101$ & 0.02646714 & 0.00910203 & 0.00200123 & 0.02490986 & 0.01000296 & 0.02565261
 & 0.02395322 & 0.00291095\\
 & $110$ & 0.03841534 & 0.00460422 & 0.00324304 & 0.01621767 & 0.03833187 & 0.00392893
 & 0.00501755 & 0.01524138\\
 & $111$ & 0.01084723 & 0.03217232 & 0.01762114 & 0.00183957 & 0.03226824 & 0.00999257
 & 0.00168794 & 0.01857098\\
\end{tabular}
\end{ruledtabular}}
\end{table*}

\subsection{MLNS estimate for the data from\texorpdfstring{~\cite{Cao2022}}{C}}\label{a:MLNS_Cao}

We now consider a count dataset from the experiment described in~\cite{Cao2022} involving a $4$-photon GHZ state (refer to Table I in Appendix A of~\cite{Cao2022} for the count data). It is presented in this paper in Table~\ref{tab:RawData_Cao}. Similar to the preceding subsection, we first process the four-photon data to obtain the three-party data. A three-photon GHZ state is obtained from a four-photon GHZ state by projecting the fourth party over the $\ket{+}$ basis and conditioning on the ``$+$" outcome. And so in Table~\ref{tab:RawData_Cao}, the data corresponding to only those measurement-outcome combinations in which the fourth photon is measured in a Pauli $\mathsf X$ with the resulting outcome ``$+$" are retained. Also, for the rows in Table~\ref{tab:RawData_Cao} involving the measurement $\mathsf{\frac{Z-X}{\sqrt 2}}$ for the second photon, for a given combination of outcomes for the first, third and fourth photon, we swap the entry for the outcome combination marked `$+$' for the second photon with the entry for the outcome combination marked `$-$' for the same photon. Again, this is necessary for the same reason as mentioned in the previous sub-section in~\ref{a:MLNS_Mao}. The original protocol in~\cite{Cao2022} has the measurements $\frac{\mathsf{X+Z}}{\sqrt{2}}$ and $\frac{\mathsf{X-Z}}{\sqrt{2}}$ for the second photon. However, in the count dataset from~\cite{Cao2022} the counts are reported for the measurements $\frac{\mathsf{Z+X}}{\sqrt{2}}$ and $\frac{\mathsf{Z-X}}{\sqrt{2}}$ for the second photon.

Next, we ignore the rows corresponding to the measurements $\mathsf{XXXX}$, $\mathsf{ZZZZ}$ and $\mathsf{ZZZX}$ because they correspond to settings combinations (for the first three parties) that do not belong to the original three-party protocol. In the original protocol parties $\mathsf{A,C}$ perform measurements $(\mathsf{Z,X})$ for the settings choices $(0,1)$ and party $\mathsf{B}$ performs measurements $\left(\mathsf{\frac{X+Z}{\sqrt 2},\frac{X-Z}{\sqrt 2}}\right)$ for the settings choices $(0,1)$. For the remaining rows we see that they correspond to some (though not all) measurement settings combinations for the first three parties (after ignoring the fourth measurement $\mathsf{X}$) that belong to the original three-party protocol. For instance, the first row in Table~\ref{tab:RawData_Cao} with measurement combination $\mathsf{X\frac{Z+X}{\sqrt 2}XX}$ represents the settings combination $101$, the second row represents $111$, the third row represents $001$ and so on. Finally, the entries shaded in grey and blue in Table~\ref{tab:RawData_Cao} are not retained in the three-party count data.

Now, notice that we do not have count data for the settings combinations $000$ and $010$. If we revisit the linear witness in~\eqref{eq:Cao}, all correlator terms (marginal as well as full) except $\mathcal{M}_{\mathsf{AC}}^{00}$ can be shown to depend on setting combinations that are directly available from the raw count data in Table~\ref{tab:RawData_Cao}. Nonetheless, data relevant to $\mathcal{M}_{\mathsf{AC}}^{00}$ and the two missing settings combinations $000$ and $010$ is obtained from the row shaded in blue in Table~\ref{tab:RawData_Cao} which represents the data for measurement combination $\mathsf{ZZZX}$. According to the three-party protocol, while $\mathsf Z$ is a valid measurement for both $\mathsf{A}\text{ and }\mathsf{C}$ (as it represents the setting choice $0$), it is not for party $\mathsf B$. But using the linear relations in~\eqref{eq:CorrVector} for the correlator $\mathcal{M}_{\mathsf{AC}}^{00}$, we observe that it is expressible as shown below:
\begin{equation}\label{eq:M_AC_00}
    \mathcal{M}_{\mathsf{AC}}^{00} = \sum_{a,b,c}(-1)^{a\oplus c}P(abc\lvert 0y0) = \sum_{a,c}(-1)^{a\oplus c}(P(a0c\lvert 0y0)+P(a1c\lvert 0y0)),\,\text{for }y\in\{0,1\}.
\end{equation}
The second equality in~\eqref{eq:M_AC_00} emphasises that, for any $y\in\{0,1\}$, to find $\mathcal{M}_{\mathsf{AC}}^{00}$ we need only the sums $P(a0c\lvert 0y0) + P(a1c\lvert 0y0)$ for the different combinations of the outcomes $a,c\in\{0,1\}$. By no-signalling, we can obtain these from the ``sum counts" $N(s+t+\lvert \mathsf{ZZZX})+N(s-t+\lvert \mathsf{ZZZX})$ for the four fixed choices of $s,t\in\{+,-\}$ from Table~\ref{tab:RawData_Cao}, where $N(\cdot\lvert\mathsf{ZZZX})$ denote counts for the relevant outcome combinations. Since the authors of~\cite{Cao2022} use number of experimental standard deviations for the individual correlator terms (both full and marginal) in their statistical analysis, it is our understanding that the relevant ``sum counts" for the row representing the measurement combination $\mathsf{ZZZX}$ sufficed for estimating the correlator $\mathcal{M}_{\mathsf{AC}}^{00}$. However, for our statistical analysis we require a three-party empirical trial distribution from processing the four-party count data, and so the individual count entries for all outcome combinations for those measurement combinations of the first three parties that represent settings combinations $000$ and $010$ are required. Our method for dealing with this is as follows: if $\#(abc\lvert 000)$ and $\#(abc\lvert 010)$ represent the counts for various combinations of $a,b,c\in\{0,1\}$ for the settings combinations $000$ and $010$, then we obtain the count data for these two settings combinations from the row in blue in Table~\ref{tab:RawData_Cao} using the following relations:
\begin{eqnarray}\label{eq:Rows000_010_forCaodata}
    \sum_{b}\#(0b0\lvert 000) = \sum_{b}\#(0b0\lvert 010) &=& N(++++\lvert \mathsf{ZZZX}) + N(+-++\lvert \mathsf{ZZZX}),  \nonumber \\
    \sum_{b}\#(0b1\lvert 000) = \sum_{b}\#(0b1\lvert 010) &=& N(++-+\lvert \mathsf{ZZZX}) + N(+--+\lvert \mathsf{ZZZX}),\nonumber \\
    \sum_{b}\#(1b0\lvert 000) = \sum_{b}\#(1b0\lvert 010) &=& N(-+-++\lvert \mathsf{ZZZX}) + N(--++\lvert \mathsf{ZZZX}),\nonumber \\
    \sum_{b}\#(1b1\lvert 000) = \sum_{b}\#(1b1\lvert 010) &=& N(-+-+\lvert \mathsf{ZZZX}) + N(---+\lvert \mathsf{ZZZX}).
\end{eqnarray}
Consider the first equality in~\eqref{eq:Rows000_010_forCaodata}. We arbitrarily choose values for $\#(000\lvert 000)$ and $\#(010\lvert 000)$ such that their sum is equal to $N(++++\lvert \mathsf{ZZZX}) + N(+-++\lvert \mathsf{ZZZX})$. We do likewise for $\#(000\lvert 010)$ and $\#(010\lvert 010)$. The same process is repeated for the other three equalities. Once we obtain the three-party count data, we find the corresponding empirical trial distribution which is shown in Table\ref{tab:EmpiricalDist_Cao}. Subsequent computations below are set up so as not to depend on the arbitrary partitioning of these sum quantities into two counts.

\begin{table*}[!htb]
\caption{\label{tab:RawData_Cao} Count data from~\cite{Cao2022} (see Table I in Appendix A). The columns and rows represent the outcomes and measurements, respectively. The three-party count data is obtained by ignoring the rows with the measurements $\mathsf{XXXX, ZZZZ, ZZZX}$ and the columns with the outcomes having a `$-$' in the fourth place from the left (shaded grey), followed by some other manipulations involving the blue-shaded row as explained in the subsection~\ref{a:MLNS_Cao}.}
\footnotesize
\resizebox{\columnwidth}{!}{%
\begin{ruledtabular}
\begin{tabular}{ccacacacacacacaca}
\multicolumn{1}{c}{} & \multicolumn{1}{c}{$\scaleto{++++}{4pt}$} & \multicolumn{1}{c}{$\scaleto{+++-}{4pt}$} & \multicolumn{1}{c}{$\scaleto{++-+}{4pt}$} & \multicolumn{1}{c}{$\scaleto{++--}{4pt}$} & \multicolumn{1}{c}{$\scaleto{+-++}{4pt}$} & \multicolumn{1}{c}{$\scaleto{+-+-}{4pt}$} & \multicolumn{1}{c}{$\scaleto{+--+}{4pt}$} & \multicolumn{1}{c}{$\scaleto{+---}{4pt}$} &
\multicolumn{1}{c}{$\scaleto{-+++}{4pt}$} & \multicolumn{1}{c}{$\scaleto{-++-}{4pt}$} & \multicolumn{1}{c}{$\scaleto{-+-+}{4pt}$} & \multicolumn{1}{c}{$\scaleto{-+--}{4pt}$} & \multicolumn{1}{c}{$\scaleto{--++}{4pt}$} & \multicolumn{1}{c}{$\scaleto{--+-}{4pt}$} & \multicolumn{1}{c}{$\scaleto{---+}{4pt}$} & \multicolumn{1}{c}{$\scaleto{----}{4pt}$} \\
\cline{2-17} $\scaleto{\mathsf{X}\frac{\mathsf{Z}+\mathsf{X}}{\sqrt{2}}\mathsf{X}\mathsf{X}}{8pt}$ & 439 & 72 & 84 & 414 & 67 & 516 & 412 & 69 & 94 & 374 & 402 & 74 & 433 & 83 & 43 & 389 \\ $\scaleto{\mathsf{X}\frac{\mathsf{Z}-\mathsf{X}}{\sqrt{2}}\mathsf{X}\mathsf{X}}{8pt}$ & 62 & 372 & 376 & 73 & 400 & 77 & 71 & 413 & 356 & 59 & 79 & 371 & 67 & 454 & 354 & 65 \\  $\scaleto{\mathsf{Z}\frac{\mathsf{Z}+\mathsf{X}}{\sqrt{2}}\mathsf{X}\mathsf{X}}{8pt}$ & 390 & 371 & 395 & 336 & 78 & 78 & 81 & 66 & 53 & 69 & 71 & 56 & 364 & 372 & 351 & 371 \\ $\scaleto{\mathsf{Z}\frac{\mathsf{Z}-\mathsf{X}}{\sqrt{2}}\mathsf{X}\mathsf{X}}{8pt}$ & 369 & 366 & 411 & 384 & 87 & 70 & 75 & 97 & 67 & 59 & 56 & 53 & 351 & 367 & 337 & 359 \\ \rowcolor{grey} $\scaleto{\mathsf{X}\mathsf{X}\mathsf{X}\mathsf{X}}{4pt}$ & 450 & 5 & 11 & 451 & 9 & 513 & 531 & 7 & 4 & 430 & 428 & 9 & 499 & 5 & 9 & 461\\ \rowcolor{grey} $\scaleto{\mathsf{Z}\mathsf{Z}\mathsf{Z}\mathsf{Z}}{4pt}$ & 8552 & 16 & 13 & 0 & 9 & 11 & 14 & 18 & 15 & 19 & 11 & 13 & 0 & 19 & 20 & 8311 \\ \rowcolor{blue!20} $\scaleto{\mathsf{Z}\mathsf{Z}\mathsf{Z}\mathsf{X}}{4pt}$ & 5229 & 4651 & 7 & 6 & 19 & 22 & 20 & 24 & 19 & 21 & 28 & 20 & 11 & 10 & 4426 & 4861 \\ $\scaleto{\mathsf{X}\frac{\mathsf{Z}+\mathsf{X}}{\sqrt{2}}\mathsf{Z}\mathsf{X}}{8pt}$ & 403 & 406 & 72 & 83 & 73 & 71 & 412 & 427 & 443 & 429 & 74 & 66 & 68 & 81 & 438 & 422\\ $\scaleto{\mathsf{X}\frac{\mathsf{Z}-\mathsf{X}}{\sqrt{2}}\mathsf{Z}\mathsf{X}}{8pt}$ & 422 & 443 & 85 & 66 & 66 & 62 & 424 & 400 & 415 & 420 & 76 & 82 & 62 & 71 & 394 & 422 \\
\end{tabular}
\end{ruledtabular}}
\end{table*}

\begin{table*}[!htb]
\caption{\label{tab:EmpiricalDist_Cao} Empirical trial distribution $f_{\mathsf{emp}}(abcxyz)$ for the count data in Table~\ref{tab:RawData_Cao} assuming a uniform settings distribution. Some frequencies in the $000$ and $010$ rows are chosen arbitrarily though they must match the condition \eqref{eq:Rows000_010_forCaodata} as discussed in the text.}
\resizebox{0.95\textwidth}{!}{%
\begin{ruledtabular}
\begin{tabular}{rrcccccccc}
\multicolumn{1}{r}{} & \multicolumn{1}{r}{} & \multicolumn{8}{c}{$abc$}\\
\cline{2-10}
\multicolumn{1}{r}{} & \multicolumn{1}{r}{} & \multicolumn{1}{c}{$000$} & \multicolumn{1}{c}{$001$} & \multicolumn{1}{c}{$010$} & \multicolumn{1}{c}{$011$} & \multicolumn{1}{c}{$100$} & \multicolumn{1}{c}{$101$} & \multicolumn{1}{c}{$110$} & \multicolumn{1}{c}{$111$} \\
\cline{2-10}
\multirow{8}{*}{\rotatebox[origin=c]{90}{$xyz$}} & $000$ & 0.03355877 & 0.00017932 & 0.03366124 & 0.00016651 & 0.00019213 & 0.0284481 & 0.00019213 & 0.0286018\\
 & $001$ & 0.02734156 & 0.02769209 & 0.00546831 & 0.00567863 & 0.00371565 & 0.00497757 & 0.02551879 & 0.0246074\\
 & $010$ & 0.03366124 & 0.00016651 & 0.03355877 & 0.00017932 & 0.00019213 & 0.0286018 & 0.00019213 & 0.0284481\\
 & $011$ & 0.00620365 & 0.00534797 & 0.02631204 & 0.0293069 & 0.02502852 & 0.02403023 & 0.00477752 & 0.00399315\\
 & $100$ & 0.02540343 & 0.00453858 & 0.00460161 & 0.02597075 & 0.02792486 & 0.00466465 & 0.00428643 & 0.02760968\\
 & $101$ & 0.02779889 & 0.00531915 & 0.00424265 & 0.02608916 & 0.00595238 & 0.02545593 & 0.02741895 & 0.0027229\\
 & $110$ & 0.00424383 & 0.02726337 & 0.02713477 & 0.00546553 & 0.00398663 & 0.02533436 & 0.02668467 & 0.00488683\\
 & $111$ & 0.02832861 & 0.00502833 & 0.00439093 & 0.0266289 & 0.00474504 & 0.02507082 & 0.02521246 & 0.0055949\\
\end{tabular}
\end{ruledtabular}}
\end{table*}
\begin{table*}[!htb]
\caption{\label{tab:MLNSest_Cao}The MLNS estimate for the empirical trial distribution in Table~\ref{tab:EmpiricalDist_Cao} assuming a uniform settings distribution.}
\resizebox{0.95\textwidth}{!}{%
\begin{ruledtabular}
\begin{tabular}{rrcccccccc}
\multicolumn{1}{r}{} & \multicolumn{1}{r}{} & \multicolumn{8}{c}{$abc$}\\
\cline{2-10}
\multicolumn{1}{r}{} & \multicolumn{1}{r}{} & \multicolumn{1}{c}{$000$} & \multicolumn{1}{c}{$001$} & \multicolumn{1}{c}{$010$} & \multicolumn{1}{c}{$011$} & \multicolumn{1}{c}{$100$} & \multicolumn{1}{c}{$101$} & \multicolumn{1}{c}{$110$} & \multicolumn{1}{c}{$111$} \\
\cline{2-10}
\multirow{8}{*}{\rotatebox[origin=c]{90}{$xyz$}} & $000$ & 0.05526551 & 0.00003347 & 0.00981915 & 0.00036254 & 0.0002421 & 0.00877659 & 0.00006518 & 0.05043548\\
 & $001$ & 0.02766765 & 0.02763136 & 0.00485753 & 0.00532414 & 0.00412317 & 0.00489547 & 0.02624344 & 0.02425723\\
 & $010$ & 0.01004867 & 0.00034732 & 0.05503599 & 0.00004868 & 0.00005569 & 0.05011917 & 0.0002516 & 0.00909291\\
 & $011$ & 0.00578076 & 0.0046152 & 0.02674442 & 0.02834029 & 0.02540394 & 0.02477091 & 0.00496266 & 0.0043818\\
 & $100$ & 0.02719486 & 0.00457093 & 0.00523716 & 0.02580832 & 0.02831269 & 0.00423915 & 0.00464719 & 0.02498969\\
 & $101$ & 0.02637421 & 0.00539159 & 0.0043694 & 0.02667608 & 0.0054166 & 0.02713524 & 0.02673157 & 0.0029053\\
 & $110$ & 0.00527976 & 0.0257657 & 0.02715226 & 0.00461355 & 0.0048246 & 0.02470077 & 0.02813528 & 0.00452807\\
 & $111$ & 0.02623523 & 0.00481023 & 0.00450838 & 0.02725743 & 0.00494948 & 0.02457589 & 0.0271987 & 0.00546465\\
\end{tabular}
\end{ruledtabular}}
\end{table*}
Next, we use the maximum-likelihood procedure to obtain a no-signalling estimate for the empirical trial distribution that conforms to the fixed and known settings distribution. However, as a consequence of the way we have handled the missing data issue, we modify the maximum-likelihood procedure in~\eqref{eq:MLNSestimate} as follows: Let $\mathcal S$ denote the settings combinations $\{0,1\}^{3}\setminus \{(0,0,0),(0,1,0)\}$. Assuming $n$ iid trials, given that the true distribution is $\mathbb Q$, the likelihood of the trial data with empirical frequencies $\{f_{\mathsf{emp}}(abcxyz)\}$ is proportional to the expression in~\eqref{eq:Cao_likelihood_expr}. Notice that for the settings combinations $(x,y,z)$ belonging to the set $\mathcal S'=\{(0,0,0),(0,1,0)\}$ we consider the sum of the probabilities $\sum_{b}Q(abcxyz)$ and raise it to the power of the corresponding sum of the frequencies $n\sum_{b}f_{\mathsf{emp}}(abcxyz)$ for all combinations of outcomes $a,c$. This is to ensure our maximum likelihood estimate $Q$ does not depend on our arbitrary choices of $f_{\mathsf{emp}}(a0cxyz)$ and $f_{\mathsf{emp}}(a1cxyz)$, chosen only to satisfy the condition that their sum aligns with the corresponding sum on the right-hand side of \eqref{eq:Rows000_010_forCaodata}.
\begin{eqnarray}\label{eq:Cao_likelihood_expr}
\prod_{(x,y,z)\in\mathcal{S}}\prod_{a,b,c}Q(abcxyz)^{nf_{\mathsf{emp}}(abcxyz)}\prod_{(x,y,z)\in\mathcal{S}'}\prod_{a,c}\left[\sum_{b}Q(abcxyz)\right]^{n\sum_{b}f_{\mathsf{emp}}(abcxyz)}
\end{eqnarray}

We then maximise the expression obtained by taking the base-2 logarithm of~\eqref{eq:Cao_likelihood_expr} (where we ignore the constant factor $n$) as shown below in~\eqref{eq:MLNSestimate_Cao}. The constraints of the maximisation involve the no-signalling conditions and conformity to the fixed settings distribution.

\begin{eqnarray}\label{eq:MLNSestimate_Cao}
    \max_{\mathbb Q}\,\,&&\sum_{(x,y,z)\in\mathcal{S}}\sum_{a,b,c}f_{\mathsf{emp}}(abcxyz)\log_{2}Q(abcxyz)+\sum_{(x,y,z)\in\mathcal{S}'}\sum_{a,c}\sum_{b}f_{\mathsf{emp}}(abcxyz)\log_{2}\sum_{b}Q(abcxyz)\nonumber \\
\text{subject to}\,\, &&\sum_{a}Q(abc\lvert xyz)=\sum_{a}Q(abc\lvert x'yz),\,\forall b,c,x,x',y,z\nonumber \\
                      &&\sum_{b}Q(abc\lvert xyz)=\sum_{b}Q(abc\lvert xy'z),\,\forall a,c,x,y,y',z\nonumber \\
                      &&\sum_{c}Q(abc\lvert xyz)=\sum_{c}Q(abc\lvert xyz'),\,\forall a,b,x,y,z,z'\nonumber \\     &&\sum_{a,b,c}Q(abcxyz)=S(xyz),\,\forall x,y,z.
\end{eqnarray}
The objective quantity in~\eqref{eq:MLNSestimate_Cao} is a concave function in $\mathbb Q$, the proof of which is shown below, and hence the routine can be implemented using any standard convex optimisation programme. While this procedure returns estimates for $Q(a0cxyz)$ and $Q(a1cxyz)$ for the missing settings $000$ and $010$, it is unclear to what degree these estimates can be considered to reflect the actually recorded data in the experiment (as opposed to the sum of these quantities whose relation to the data is more well motivated). Thus in the main text, we choose to constrain the test factors to map $a0cxyz$ and $a1cxyz$ counts the same way on these settings.
\begin{prop}
    For the set $\mathcal{S}\coloneqq\{0,1\}^{3}\setminus\{(0,0,0),(0,1,0)\}$ and $a,b,c\in\{0,1\}$ the functional defined as:
\begin{equation*}
    G(\mathbb Q) \coloneqq \sum_{(x,y,z)\in\mathcal{S}}\sum_{a,b,c}f_{\mathsf{emp}}(abcxyz)\log_{2}Q(abcxyz)+\sum_{(x,y,z)\in\mathcal{S}'}\sum_{a,c}\sum_{b}f_{\mathsf{emp}}(abcxyz)\log_{2}\sum_{b}Q(abcxyz)
\end{equation*}
    is concave in $\mathbb{Q}\coloneqq\{Q(abcxyz)\}$.  
\begin{proof}
    To prove concavity of $G(\mathbb Q)$ we need to show that $G(\mathbb Q)\ge \lambda G(\mathbb{Q}')+(1-\lambda)G(\mathbb{Q}'')$ whenever $\mathbb{Q}=\lambda\mathbb{Q}'+(1-\lambda)\mathbb{Q}''$ for $0\le\lambda\le 1$. In the expressions that follow, we use the concise notations $\sum_{\mathcal S}$ and $\sum_{\mathcal{S}'}$ to mean, respectively, the sum over the settings combinations $(x,y,z)$ belonging to the sets $\mathcal{S}$ and $\mathcal{S}'$. We then can use the concavity of $\log$ and manipulation of sums to write
    \begin{eqnarray*}
    &&G(\lambda\mathbb{Q}'+(1-\lambda)\mathbb{Q}'')\\
    =&&\sum_{\mathcal{S}}\sum_{a,b,c}f_{\mathsf{emp}}(abcxyz)\log_{2}\big[\lambda Q'(abcxyz) + (1-\lambda)Q''(abcxyz)\big] + \\
    &&\sum_{\mathcal{S}'}\sum_{a,c}\sum_{b}f_{\mathsf{emp}}(abcxyz)\log_{2}\big[\lambda \sum_{b}Q'(abcxyz)+(1-\lambda)\sum_{b}Q''(abcxyz)\big] \\
    \ge&& \sum_{\mathcal{S}}\sum_{a,b,c}f_{\mathsf{emp}}(abcxyz)\big[\lambda\log_{2} Q'(abcxyz)+(1-\lambda)\log_2 Q''(abcxyz)\big] + \\
    &&\sum_{\mathcal{S}'}\sum_{a,c}\sum_{b}f_{\mathsf{emp}}(abcxyz)\big[\lambda\log_{2}\sum_{b}Q'(abcxyz) + (1-\lambda)\log_{2}\sum_{b}Q''(abcxyz)\big]\\
    =&& \lambda\Big(\sum_{\mathcal{S}}\sum_{a,b,c}f_{\mathsf{emp}}(abcxyz)\log_{2} Q'(abcxyz) + \sum_{\mathcal{S}'}\sum_{a,c}\sum_{b}f_{\mathsf{emp}}(abcxyz)\log_{2}\sum_{b}Q'(abcxyz)\Big)\\
    +&& (1-\lambda)\Big(\sum_{\mathcal{S}}\sum_{a,b,c}f_{\mathsf{emp}}(abcxyz)\log_{2} Q''(abcxyz) + \sum_{\mathcal{S}'}\sum_{a,c}\sum_{b}f_{\mathsf{emp}}(abcxyz)\log_{2}\sum_{b}Q''(abcxyz)\Big) \\
    =&& \lambda G(\mathbb{Q}') + (1-\lambda)G(\mathbb{Q}'').
\end{eqnarray*}
\end{proof}
\end{prop}
The MLNS estimate for the empirical trial distribution in Table~\ref{tab:EmpiricalDist_Cao} is then found by implementing the maximisation routine in~\eqref{eq:MLNSestimate_Cao} and is as shown in Table~\ref{tab:MLNSest_Cao}.

\section{Vertex Enumeration}\label{a:VertEnum}

In this work we perform several vertex enumeration routines using the software $\mathsf{polymake}$~\cite{Gawrilow2000}. The first such task is that of enumerating the vertices of the three-party no-signalling polytope for binary settings and outcomes. The 48 no-signalling conditions in the linear relations expressed in~\eqref{eq:NS_overA}-~\eqref{eq:NS_overC} and the 8 normalisation conditions together constitute a set of hyperplanes, and the 64 non-negativity conditions constitute a set of closed half-spaces. The set of hyperplanes are the equality conditions expressible in a matrix format as $(\mathbf{H}_{\mathrm{eq}})_{56\times 64}\mathbf{P}_{64\times 1}=\mathbf{b}_{56\times 1}$, and the closed half-spaces are represented by the inequalities $(\mathbf{H}_{\mathrm{ineq}})_{64\times 64}\mathbf{P}_{64\times 1}\ge \mathbf{0}_{64\times 1}$. The no-signalling polytope $\Xi_{\mathsf{ns}}$ can then be represented as an intersection of finite number of closed half-spaces and hyperplanes as:
\begin{equation}
    \Xi_{\mathsf{ns}} \coloneqq \left\{\mathbf{P}\in\mathbb{R}^{64}\colon \mathbf{H}_{\mathrm{eq}}\mathbf{P}=\mathbf{b},\mathbf{H}_{\mathrm{ineq}}\mathbf{P}\ge \mathbf{0}\right\}.
\end{equation}
Note that a subset of the total $48$ no-signalling conditions is redundant, as these can be derived from a combination of normalisation conditions and the non-redundant no-signalling conditions. However, for the purpose of generating the vertices of $\Xi_{\mathsf{ns}}$ from its hyperplane representation using the $\mathsf{polymake}$ program, including these redundant no-signalling conditions does not pose any issue. There are 53856 vertices of the polytope $\Xi_{\mathsf{ns}}$, and they can be classified into 46 equivalence classes~\cite{ExtrCorrTripartite_Pironio}. 

Next, to approximate the set $\Delta_{\mathsf{lo2sr}}$ with a polytope, we consider the intersection of $\Xi_{\mathsf{ns}}$ with the inequality in~\eqref{eq:Mao}. Another polytope approximating $\Delta_{\mathsf{lo2sr}}$ is the one obtained by the intersection of $\Xi_{\mathsf{ns}}$ with the inequality in~\eqref{eq:Cao}. We can express both inequalities~\eqref{eq:Mao} and~\eqref{eq:Cao} as $\mathbf{B}\cdot\mathbf{P}\le\beta$, where $\{B(abcxyz)\}\eqqcolon\mathbf{B}\in\mathbb{R}^{64}$ is a Bell vector. As mentioned earlier in Section~\ref{s:Bell_GTNL_witness}, the Bell vector is not unique due to the no-signalling condition: the marginal correlators in the inequalities remain invariant under the settings choices of the party (or parties) not, so that a correlator such as for example $\mathcal{M}_{\mathsf{AB}}^{00}$ can be represented multiple different ways as a linear combination of $P(abc|00z)$ terms using different choices of $z$. We express the inequalities in~\eqref{eq:Mao} and~\eqref{eq:Cao} as $\mathbf{B}_{(1)}\cdot\mathbf{P}\le 4$ and $\mathbf{B}_{(2)}\cdot\mathbf{P}\le 8$, where the Bell vectors $\mathbf{B}_{(1)}$ and $\mathbf{B}_{(2)}$ are as shown in tabulated form in Tables~\ref{tab:B1} and~\ref{tab:B2}. The sets $\Xi_{(1)}\coloneqq\{\mathbf{P}\in\Xi_{\mathsf{ns}}\colon\mathbf{B}_{(1)}\cdot\mathbf{P}\le 4\}$ and $\Xi_{(2)}\coloneqq\{\mathbf{P}\in\Xi_{\mathsf{ns}}\colon\mathbf{B}_{(2)}\cdot\mathbf{P}\le 8\}$ are the respective intersection of $\Xi_{\mathsf{ns}}$ with the closed half-spaces formed by the inequalities $\mathbf{B}_{(1)}\cdot\mathbf{P}\le 4$ and $\mathbf{B}_{(2)}\cdot\mathbf{P}\le 8$, and result in polytopes with $56767$ and $57283$ extreme points, respectively.
\begin{table}[!htb]
\begin{minipage}{0.48\textwidth}
\caption{\label{tab:B1} Tabular representation of a Bell vector $\mathbf{B}_{(1)}$ corresponding to the inequality in~\eqref{eq:Mao}.}
{\renewcommand{\arraystretch}{1}%
\begin{ruledtabular}
\begin{tabular}{rrcccccccc}
\multicolumn{1}{r}{} & \multicolumn{1}{r}{} & \multicolumn{8}{c}{$abc$}\\
\cline{2-10}
\multicolumn{1}{r}{} & \multicolumn{1}{r}{} & \multicolumn{1}{c}{$000$} & \multicolumn{1}{c}{$001$} & \multicolumn{1}{c}{$010$} & \multicolumn{1}{c}{$011$} & \multicolumn{1}{c}{$100$} & \multicolumn{1}{c}{$101$} & \multicolumn{1}{c}{$110$} & \multicolumn{1}{c}{$111$} \\
\cline{2-10}
\multirow{8}{*}{\rotatebox[origin=c]{90}{$xyz$}} & $000$ & 3 & -1 & 1 & -3 & -3 & 1 & -1 & 3 \\
 & $001$ & 0 & 0 & 0 & 0 & 0 & 0 & 0 & 0\\
 & $010$ & 1 & 1 & -1 & -1 & -1 & -1 & 1 & 1 \\
 & $011$ & 0 & 0 & 0 & 0 & 0 & 0 & 0 & 0 \\
 & $100$ & 0 & 0 & 0 & 0 & 0 & 0 & 0 & 0 \\
 & $101$ & 1 & -1 & -1 & 1 & -1 & 1 & 1 & -1 \\
 & $110$ & 0 & 0 & 0 & 0 & 0 & 0 & 0 & 0 \\
 & $111$ & -1 & 1 & 1 & -1 & 1 & -1 & -1 &  1
\end{tabular}
\end{ruledtabular}}
\end{minipage}
\hfill
\begin{minipage}{0.48\textwidth}
\caption{\label{tab:B2} Tabular representation of a Bell vector $\mathbf{B}_{(2)}$ corresponding to the inequality in~\eqref{eq:Cao}.}
{\renewcommand{\arraystretch}{1}%
\begin{ruledtabular}
\begin{tabular}{rrcccccccc}
\multicolumn{1}{r}{} & \multicolumn{1}{r}{} & \multicolumn{8}{c}{$abc$}\\
\cline{2-10}
\multicolumn{1}{r}{} & \multicolumn{1}{r}{} & \multicolumn{1}{c}{$000$} & \multicolumn{1}{c}{$001$} & \multicolumn{1}{c}{$010$} & \multicolumn{1}{c}{$011$} & \multicolumn{1}{c}{$100$} & \multicolumn{1}{c}{$101$} & \multicolumn{1}{c}{$110$} & \multicolumn{1}{c}{$111$} \\
\cline{2-10}
\multirow{8}{*}{\rotatebox[origin=c]{90}{$xyz$}} & $000$ & 4 & -4 &  4 & -4 & -4 &  4 & -4 & 4 \\
 & $001$ & 1 & 1 & -1 & -1 & -1 & -1 & 1 &  1\\
 & $010$ & -1 & 1 & 1 & -1 & -1 & 1 & 1 & -1 \\
 & $011$ & -1 & -1 & 1 & 1 & 1 & 1 & -1 & -1 \\
 & $100$ & 1 & -1 & -1 & 1 & 1 & -1 & -1 &  1 \\
 & $101$ & 2 & -2 & -2 & 2 & -2 & 2 & 2 & -2 \\
 & $110$ & 0 & 0 & 0 & 0 & 0 & 0 & 0 & 0 \\
 & $111$ & 2 & -2 & -2 & 2 & -2 & 2 & 2 & -2
\end{tabular}
\end{ruledtabular}}
\end{minipage}
\end{table}
The vertex enumeration routines that we perform in our work are summarised in Table~\ref{tab:VertEnum}. As mentioned in the main text, the number of extreme points of the approximating polytopes correspond to the number of constraints in the respective test factor optimisation routines and, as explained earlier, we were able to reduce that number by restricting the optimisation feasibility region by requiring the condition $\mathsf{E}[F]\le 1$ for only those no-signalling behaviours that saturate the inequality intersecting $\Xi_{\mathsf{ns}}$\PB{,} which brings us to the vertex enumeration for the intersection of the no-signalling polytope with a Bell hyperplane: $\Xi_{(1)}'\coloneqq\{\mathbf{P}\in\Xi_{\mathsf{ns}}\colon\mathbf{B}_{(1)}\cdot\mathbf{P}=4\}$ and $\Xi_{(2)}'\coloneqq\{\mathbf{P}\in\Xi_{\mathsf{ns}}\colon\mathbf{B}_{(2)}\cdot\mathbf{P}=8\}$. The polytopes $\Xi_{(1)}'$ and $\Xi_{(2)}'$ have $3200$ and $3664$ extreme points, respectively. There are interesting features of the set $\mathsf{Ext}(\Xi_{\mathsf{ns}})$ with regards to the two linear witnesses. For instance, there is only one behaviour in $\mathsf{Ext}(\Xi_{\mathsf{ns}})$, denoted as $\mathbf{P}_{(1)}$ in~\eqref{eq:ExtBoxMaxViolMao}, that maximally violates $\mathbf{B}_{(1)}\cdot\mathbf{P}\le 4$, achieving the value $6$, and belongs to the equivalence class 8 (see Table~1 in~\cite{ExtrCorrTripartite_Pironio}) of the extreme points of $\Xi_{\mathsf{ns}}$. A relabelled version $\mathbf{P}_{(2)}$ of this behaviour (and hence belonging to the same equivalence class) maximally violates $\mathbf{B}_{(2)}\cdot\mathbf{P}\le 8$ achieving the value $12$.

\begin{eqnarray}\label{eq:ExtBoxMaxViolMao}
    \mathbf{P}_{(1)}\coloneqq\begin{cases}
        P(0bc\lvert 0yz) = \frac{1}{4}\delta_{b,0}(\delta_{c,0}+\delta_{c,z})\\
        P(0bc\lvert 1yz) = \frac{1}{4}\delta_{b\oplus c,yz}\\
        P(1bc\lvert 0yz) = \frac{1}{4}\delta_{b,1}(\delta_{c,1} + \delta_{c,z\oplus 1})\\
        P(1bc\lvert 1yz) = \frac{1}{4}\delta_{b\oplus c,yz\oplus z}
    \end{cases},\,\mathbf{P}_{(2)}\coloneqq\begin{cases}
        P(0bc\lvert 0yz) = \frac{1}{4}\delta_{b,y}(\delta_{c,0}+\delta_{c,z})\\
        P(0bc\lvert 1yz) = \frac{1}{4}\delta_{b\oplus c,yz\oplus y} \\
        P(1bc\lvert 0yz) = \frac{1}{4}\delta_{b,y\oplus 1}(\delta_{c,1}+\delta_{c,z\oplus 1}) \\
        P(1bc\lvert 1yz) = \frac{1}{4}\delta_{b\oplus c,yz\oplus z\oplus y} 
    \end{cases},\forall b,c,y,z 
\end{eqnarray}

\begin{table*}[!htb]
\caption{\label{tab:VertEnum}Vertex Enumeration.}
{\renewcommand{\arraystretch}{1.1}%
\begin{ruledtabular}
\begin{tabular}{cccc}
  \multicolumn{1}{c}{} & Polytope  & Number of Vertices & Remarks \\
   \hline
 (1) & $\Xi_{\mathsf{ns}}$  & 53856 & No-signalling polytope for (3,2,2) scenario \\
 (2) & $\Xi_{(1)}'\coloneqq\{\mathbf{P}\in\Xi_{\mathsf{ns}}\colon\mathbf{B}_{(1)}\cdot\mathbf{P}=4\}$ & 3200 & Intersection of $\Xi_{\mathsf{ns}}$ and a hyperplane \\
 (3) & $\Xi_{(1)}\coloneqq\{\mathbf{P}\in\Xi_{\mathsf{ns}}\colon\mathbf{B}_{(1)}\cdot\mathbf{P}\le 4\}$ & 56767 & Intersection of $\Xi_{\mathsf{ns}}$ and a closed half-space\\
 (4) & $\Xi_{(2)}'\coloneqq\{\mathbf{P}\in\Xi_{\mathsf{ns}}\colon\mathbf{B}_{(2)}\cdot\mathbf{P}=4\}$ & 3664 & Intersection of $\Xi_{\mathsf{ns}}$ and a hyperplane \\
 (5) & $\Xi_{(2)}\coloneqq\{\mathbf{P}\in\Xi_{\mathsf{ns}}\colon\mathbf{B}_{(2)}\cdot\mathbf{P}\le 8\}$ & 57283 & Intersection of $\Xi_{\mathsf{ns}}$ and a closed half-space \\
\end{tabular}
\end{ruledtabular}}
\end{table*}

Table~\ref{tab:ExtPoints_NSEqClass} categorises the candidates of the set $\mathsf{Ext}(\Xi_{\mathsf{ns}})$, which contains the extreme points of $\Xi_{\mathsf{ns}}$, into various equivalence classes based on their saturation and violation of the two DI witnesses. For example, there are $8$ entries from equivalence class $2$, the ``Popescu-Rohrlich (PR)" class of behaviours~\cite{ExtrCorrTripartite_Pironio}, that saturate equation~\eqref{eq:Mao}, and $4$ entries from the same class that saturate equation~\eqref{eq:Cao}. Additionally, there are $32$ entries from equivalence class $25$, the ``Guess Your Neighbour's Input (GYNI)" class of behaviours~\cite{ExtrCorrTripartite_Pironio}, that saturate equation~\eqref{eq:Mao}, and $24$ entries from the same class that saturate equation~\eqref{eq:Cao}. Interestingly, there are no entries from equivalence class $29$ that saturate either equation~\eqref{eq:Mao} or equation~\eqref{eq:Cao}, despite this class also being categorised as a GYNI class of behaviours~\cite{ExtrCorrTripartite_Pironio}. However, equivalence class $29$ does have $8$ entries that violate equation~\eqref{eq:Mao} and $8$ (possibly different) entries that violate equation~\eqref{eq:Cao}. As demonstrated in Table~\ref{tab:ExtPoints_NSEqClass}, there are numerous extreme points of $\Xi_{\mathsf{ns}}$ belonging to other equivalence classes that violate the DI witnesses in equations~\eqref{eq:Mao} and~\eqref{eq:Cao}, and these classes were not explored in depth in~\cite{ExtrCorrTripartite_Pironio}.

\begin{table*}[!htb]
\caption{\label{tab:ExtPoints_NSEqClass} Extreme points of $\Xi_{\mathsf{ns}}$ from the different equivalence classes described in \cite{ExtrCorrTripartite_Pironio} that saturate and violate the linear witnesses.}
    \begin{ruledtabular}
        \begin{tabular}{cccccc}
           Eq.~Class  & \# saturating~\eqref{eq:Mao} & \# violating ~\eqref{eq:Mao} & \# saturating~\eqref{eq:Cao} & \# violating~\eqref{eq:Cao} & Remarks \\
           \hline
           1 & 16 & - & 16 & - & Local Deterministic Behaviours\\
           2 & 8 & - & 4 & - & Popescu-Rohrlich Behaviours\\
           3 & 2 & - & - & - & \\
           5 & 8 & - & 8 & - & \\
           6 & 10 & - & 8 & - & \\
           8 & 8 & 1 & 8 & 1 & \\
           10 & 8 & 4 & 8 & 4 & \\
           12 & 20 & 4 & 16 & 8 & \\
           14 & 16 & - & 16 & - & \\
           15 & - & 16 & 8 & 16 & \\
           17 & 8 & 16 & - & 16 & \\
           18 & 16 & - & 8 & - & \\
           24 & 20 & 4 & 16 & 8 & \\
           25 & 32 & - & 24 & - & GYNI Behaviours \\
           26 & 16 & - & 16 & - & \\
           27 & 8 & - & - & - & \\
           28 & 16 & - & 16 & - & \\
           29 & - & 8 & - & 8 & GYNI Behaviours \\
           32 & 8 & - & - & - & \\
           40 & 4 & - & - & - & \\
           41 & 4 & - & 4 & - & \\
           42 & 8 & - & 8 & - & \\
        \end{tabular}
    \end{ruledtabular}
\end{table*}

\section{\label{a:JWPanDIWitness_Linear}Another linear DI witness for genuine tripartite nonlocality}

Here we prove that the inequality in~\eqref{eq:JWPan_ineq} used as a DI witness in~\cite{JianWeiPan2022} is linear and hence can be expressed in the form $\mathbf{B}\cdot\mathbf{P}\le\beta$. As mentioned in the main text, the inequality is a combination of two Bell games: (1) the CHSH game between parties $\mathsf A$ and $\mathsf B$ conditioned on $\mathsf C$ obtaining the outcome $c=0$ for the measurement setting $z=1$, expressed as
\begin{equation}\label{eq:CHSH_given_C}
    \mathsf{CHSH}_{z=1}^{c=0}\coloneqq\sum_{x,y}(-1)^{xy}\frac{\sum_{a,b}(-1)^{a\oplus b}P(ab0\lvert xy1)}{\sum_{a,b}P(ab0\lvert xy1)},
\end{equation}
and (2) the ``Same" game, expressed as 
\begin{equation}\label{eq:Same_game}
\mathsf{Same}\coloneqq \mathcal{M}_{\mathsf{AB}}^{02}+\mathcal{M}_{\mathsf{BC}}^{20}.
\end{equation}
The condition for perfect score for the $\mathsf{Same}$ expression requires all parties to obtain identical outcomes which can occur in one of two ways: both parties observe $0$, or both parties observe $1$. Violation of~\eqref{eq:JWPan_ineq} serves as a DI witness for genuine tripartite nonlocality in the sense that the correlations are not achievable in a scenario where every set of two parties share a (possibly generalised) nonlocal resource and all parties have access to unlimited shared randomness. Here we show that it is a linear inequality by expressing it as:
\begin{equation}\label{eq:JWPan_ineq_sumofprob}
    \sum_{s,t=0}^{1}\left[(-1)^{s\oplus t}\left\{P(st0\lvert 011)+P(st0\lvert 101)-P(st0\lvert 111)+4P(sts\lvert 020)\right\}-3^{s\oplus t}P(st0\lvert 001)\right]\le 4.
\end{equation}
We begin by writing the $\mathsf{Same}$ expression in~\eqref{eq:Same_game} as a weighted sum of probabilities using the linear relations in~\eqref{eq:CorrVector} for the marginal correlators $\mathcal{M}_{\mathsf{AB}}^{02}$ and $\mathcal{M}_{\mathsf{BC}}^{20}$, noting that no-signaling ensures that the expression for $\mathcal{M}_{\mathsf{AB}}^{02}$ (resp., $\mathcal{M}_{\mathsf{BC}}^{20}$) is valid for any choice of third measurement $z$ (resp., $x$):
\begin{eqnarray}
    \mathsf{Same} &=& \sum_{a,b,c}(-1)^{a\oplus b}P(abc\lvert 020) + \sum_{a,b,c}(-1)^{b\oplus c}P(abc\lvert 020)\label{eq:Same_expr_step1}\\
                  &=&2\left(P(000\lvert 020)-P(010\lvert 020)-P(101\lvert 020)+P(111\lvert 020)\right)\label{eq:Same_expr_step2}\\
                  &=&2\sum_{s,t}(-1)^{s\oplus t}P(sts\lvert 020).
\end{eqnarray}
The equality in~\eqref{eq:Same_expr_step2} results from expanding the two sums in~\eqref{eq:Same_expr_step1} followed by a cancellation of some terms. As an aside, notice that the $\mathsf{Same}$ expression achieves the perfect score of $2$ if $P(a=b=c\lvert 020)=P(000\lvert 020)+P(111\lvert 020)=1$, i.e., the parties should observe the same outcomes to achieve the perfect score. 

Next, using the linear relation in~\eqref{eq:CorrVector} for $\mathcal{M}_{\mathsf C}^{1}$ and the normalisation condition that for any combination of settings the probabilities $P(abc\lvert xyz)$ must sum (over the outcomes) to one, we can express the term $1+\mathcal{M}_{\mathsf C}^{1}$ appearing in the denominator of the second term on the left hand side of the inequality in~\eqref{eq:JWPan_ineq} as:
\begin{equation}\label{eq:1+M_C}
1+\mathcal{M}_{\mathsf C}^{1} = \sum_{a,b,c}P(abc\lvert xy1) + \sum_{a,b,c}(-1)^{c}P(abc\lvert xy1)=2\sum_{a,b}P(ab0\lvert xy1).
\end{equation}
The second equality in~\eqref{eq:1+M_C} follows from expanding the two sums and a cancellation of some terms. Since by no-signalling~\eqref{eq:1+M_C} holds for any combination of $x,y$, we choose $x=y=0$. And then using the expression for $\mathsf{CHSH}_{z=1}^{c=0}$ in~\eqref{eq:CHSH_given_C} we rewrite the left hand side of the inequality in~\eqref{eq:JWPan_ineq} entirely in terms of the probabilities $P(abc\lvert xyz)$ as shown below:
\begin{equation}\label{eq:JWP_expanded_1}
\mathsf{CHSH}_{z=1}^{c=0}+\frac{4\mathsf{Same}-8}{1+\mathcal{M}_{\mathsf{C}}^{1}} = \sum_{x,y}(-1)^{xy}\frac{\sum_{a,b}(-1)^{a\oplus b}P(ab0\lvert xy1)}{\sum_{a,b}P(ab0\lvert xy1)} + \frac{4\left(\sum_{s,t}(-1)^{s\oplus t}P(sts\lvert 020)-1\right)}{\sum_{a,b}P(ab0\lvert 001)} 
\end{equation}
Consider the first summation on the right hand side of the equality in~\eqref{eq:JWP_expanded_1}. The summand's denominator term $\sum_{a,b}P(ab0\lvert xy1)$ is unchanged for any combination of $x,y$ due to the no-signalling condition in~\eqref{eq:NS_over_AB} that the composite system formed by parties $\mathsf A$ and $\mathsf B$ cannot signal to $\mathsf{C}$. In other words, the outcomes observed by the party $\mathsf C$ are not influenced by the settings choices of the composite system $\mathsf{(A,B)}$. Hence, for $(x,y)\in\{(0,1),(1,0),(1,1)\}$ we have: 
\begin{equation}\label{eq:AB_NoSignal_C}
    \sum_{a,b}P(ab0\lvert xy1) = \sum_{a,b}P(ab0\lvert 001),
\end{equation}
using which we can rewrite the right hand side of~\eqref{eq:JWP_expanded_1} as:
\begin{equation}
\sum_{x,y}(-1)^{xy}\frac{\sum_{a,b}(-1)^{a\oplus b}P(ab0\lvert xy1)}{\sum_{a,b}P(ab0\lvert 001)} + \frac{4\left(\sum_{s,t}(-1)^{s\oplus t}P(sts\lvert 020)-1\right)}{\sum_{a,b}P(ab0\lvert 001)},
\end{equation}
We then substitute the expression obtained above in the inequality in~\eqref{eq:JWPan_ineq} and multiplying both sides by $\sum_{a,b}P(ab0|001)$ obtain:
\begin{equation}
    \sum_{x,y}(-1)^{xy}\sum_{a,b}(-1)^{a\oplus b}P(ab0\lvert xy1) + 4\left(\sum_{s,t}(-1)^{s\oplus t}P(sts\lvert 020) - 1\right) \le 2\sum_{a,b}P(ab0\lvert 001).
\end{equation}
For the double summation on the left hand side of the above inequality, we expand the terms over all possible values of $x,y$, and then re-express the inequality as follows:
\begin{eqnarray*}
\sum_{a,b}\big((-1)^{a\oplus b} - 2\big)P(ab0\lvert 001) + \sum_{a,b}(-1)^{a\oplus b}\big(P(ab0\lvert 011) + P(ab0\lvert 101) - P(ab0\lvert 111)\big) + 4\sum_{s,t}(-1)^{s\oplus t}P(sts\lvert 020) \le 4.
\end{eqnarray*}
Notice that $\sum_{a,b}\big((-1)^{a\oplus b} - 2\big)P(ab0\lvert 001)=-\sum_{a,b}3^{a\oplus b}P(ab0\lvert 001)$. Now for the summations over binary variables $a,b$, one can instead choose to write them as summations over the binary variables $s,t$. The above inequality can then be re-expressed as~\eqref{eq:JWPan_ineq_sumofprob}.

\end{document}